\documentclass[tradiabstract]{aa}
\usepackage{graphicx}
\usepackage{txfonts}
\usepackage{color}

\usepackage{amsmath,amsfonts,amssymb}
\usepackage{fixltx2e}
\usepackage{caption,subcaption}
\usepackage[breaklinks, colorlinks, citecolor=blue]{hyperref}
\usepackage{natbib}
\usepackage{tabularx}
\usepackage[export]{adjustbox}

\bibpunct{(}{)}{;}{a}{}{,}
\newcommand{\Planck}{{\em Planck}}
\newcommand{\Herschel}{{\em Herschel}}
\newcommand{\CORE}{{\em CORE}}
\newcommand{\XMM}{{\em XMM-Newton}}

\begin{document}

   \title{Dust in galaxy clusters: Modeling at millimeter wavelengths and impact on \Planck\ cluster cosmology}

   \author{J.-B. Melin
          \inst{1}
          \and J. G. Bartlett
           \inst{2,3}
           \and Z.-Y. Cai
           \inst{4,5}
           \and G. De Zotti
           \inst{6}
           \and J. Delabrouille
           \inst{2}
           \and M. Roman
           \inst{7}
           \and A. Bonaldi
           \inst{8}
          }

   \institute{IRFU, CEA, Universit{\'e} Paris-Saclay, F-91191 Gif-sur-Yvette, France\\
                 \email{jean-baptiste.melin@cea.fr}
         \and
                 APC, AstroParticule et Cosmologie, Universit{\'e} Paris Diderot, CNRS/IN2P3, CEA/lrfu, Observatoire de Paris, Sorbonne Paris Cit{\'e}, 10, rue Alice Domon et L{\'e}onie Duquet, F-75205 Paris Cedex 13, France
            \and
                  Jet Propulsion Laboratory, California Institute of Technology, 4800 Oak Grove Drive, Pasadena, California, USA
            \and
                  CAS Key Laboratory for Research in Galaxies and Cosmology, Department of Astronomy, University of Science and Technology of China, Hefei 230026, China
             \and
                  School of Astronomy and Space Science, University of Science and Technology of China, Hefei 230026, China
             \and
                  INAF-Osservatorio Astronomico di Padova, Vicolo dell'Osservatorio 5, I-35122 Padova, Italy
             \and
                  Laboratoire de Physique Nucl{\'e}aire et des Hautes Energies, UPMC Univ. Paris 6, UPD Univ. Paris 7, CNRS IN2P3, 4 place Jussieu 75005, Paris, France
              \and
                  SKA Organisation, Lower Withington Macclesfield, Cheshire SK11 9DL, U.K
             }

   \date{Received ...; accepted ...}

% \abstract{}{}{}{}{} 
% 5 {} token are mandatory
 
  \abstract
{We have examined dust emission in galaxy clusters at millimeter wavelengths using the \Planck\ $857 \, {\rm GHz}$ map to constrain the model based on {\em Herschel} observations that was used in studies for the Cosmic ORigins Explorer (CORE) mission concept. By stacking the emission from \Planck-detected clusters, we estimated the normalization of the infrared luminosity versus mass relation and constrained the spatial profile of the dust emission. We used this newly constrained model to simulate clusters that we inject into \Planck\ frequency maps. The comparison between clusters extracted using these gas+dust simulations and the basic gas-only simulations allows us to assess the impact of cluster dust emission on \Planck\ results. In particular, we determined the impact on cluster parameter recovery (size, flux) and on \Planck\ cluster cosmology results (survey completeness, determination of cosmological parameters). We show that dust emission has a negligible effect on the recovery of individual cluster parameters for the \Planck\ mission, but that it impacts the cluster catalog completeness, reducing the number of detections in the redshift range [0.3-0.8] by up to $\sim 9\%$. Correcting for this incompleteness in the cosmological analysis has a negligible effect on cosmological parameter measurements: in particular, it does not ease the tension between \Planck\ cluster and primary cosmic microwave background cosmologies.}


   \maketitle
%
%-------------------------------------------------------------------

\section{Introduction}
Quantifying dust emission from galaxy clusters is interesting for both astrophysical and cosmological studies.  Dust emission from member galaxies is a tracer of the star formation rate (SFR) in dense environments \citep{alberts2014, alberts2016}, and the question of intracluster dust embedded in the hot intracluster medium (ICM) concerns stellar feedback and the physical state of the ICM \citep{montier2004}.  Cluster dust emission also has potentially important ramifications for cosmology from Sunyaev-Zeldovich (SZ) cluster counts because it can contaminate the SZ signal and modify survey selection functions.    

One of the first appearances of contaminating dust emission was found by \cite{planck2013-xi} when stacking the SZ signal from central halo galaxies.  Contamination by dust emission came to dominate the SZ signal when approaching the low-mass group scale.  \citet{hurier2016} and \citet{comis2016} examine dust emission by stacking signal in the high frequency \Planck\ maps around massive clusters from the \Planck\ catalogs.  The former work separated the dust and SZ signals to conclude that the dust emission evolved with redshift and was more spatially extended than the SZ signal.  The authors of the latter work combined IRAS data with \Planck\ observations to measure dust temperature and determine dust masses in cluster systems.  These studies extend the work of \citet{montier2005} and \citet{giard2008}, who detected dust emission by stacking IRAS maps around clusters.

Because the \Planck\ beam has an angular extent similar to or larger than that of the studied clusters, these observations integrate their total emission. \citet{hurier2016} find that the dust emission could be fully accounted for by cluster member galaxies, in agreement with previous work \citep{roncarelli2010}, a conclusion further supported by the temperatures of $T\sim 20$\,K determined by \citet{comis2016} that are typical of late-type galaxies.  

Little attention has yet been given to studying the impact of dust emission on SZ cluster surveys, largely because the level of the emission relative to the SZ signal is poorly known.  The large surveys by the Atacama Cosmology Telescope \citep[ACT,][]{hasselfield2013}, the South Pole Telescope \citep[SPT,][]{bleem2015} and the \Planck\ mission \citep{planckcatalogue2016} do not model the effect of dust emission on their selection functions and photometry.  Any effect will depend in detail on the observation bands and how they are used in cluster detection.  

In this paper, we examine dust emission from massive, intermediate redshift clusters using \Planck\ observations and evaluate its impact on the \Planck\ SZ cluster selection function and photometry. The data are presented in Sect.~\ref{sec:data}.  We proceed by first establishing a baseline model (Sect. \ref{sec:dust_model}) for cluster dust emission and fit key model parameters with our \Planck\ measurements (Sect.~\ref{sec:constraints}). These parameters are the normalization of the infrared (IR) luminosity-cluster mass relation and the spatial extent of the dust emission.  We then simulate \Planck\ observations of clusters with both SZ signal and dust emission to quantify the effect of the dust emission on the \Planck\ SZ selection function and photometry (size and SZ flux) in Sect.~\ref{sec:impact}. We conclude in Sect.~\ref{sec:ccl}.
Throughout the paper, we adopt the \Planck\ $\Lambda$CDM cosmology \citep[TT,TE,EE+lowP+lensing+ext in Table 4 of][]{planckcosmo2016}: $h = H_0 / (100 \, \rm{km s^{-1} Mpc^{-1})} = 0.6774$, $\Omega_{\rm m}=1-\Omega_{\Lambda}=0.3089$, $\Omega_b=0.0485976$, $n_{\rm s}=0.9667$.

\begin{figure}
\centering
\adjincludegraphics[width=\hsize,trim={0 0 0 {.25\width}},clip]{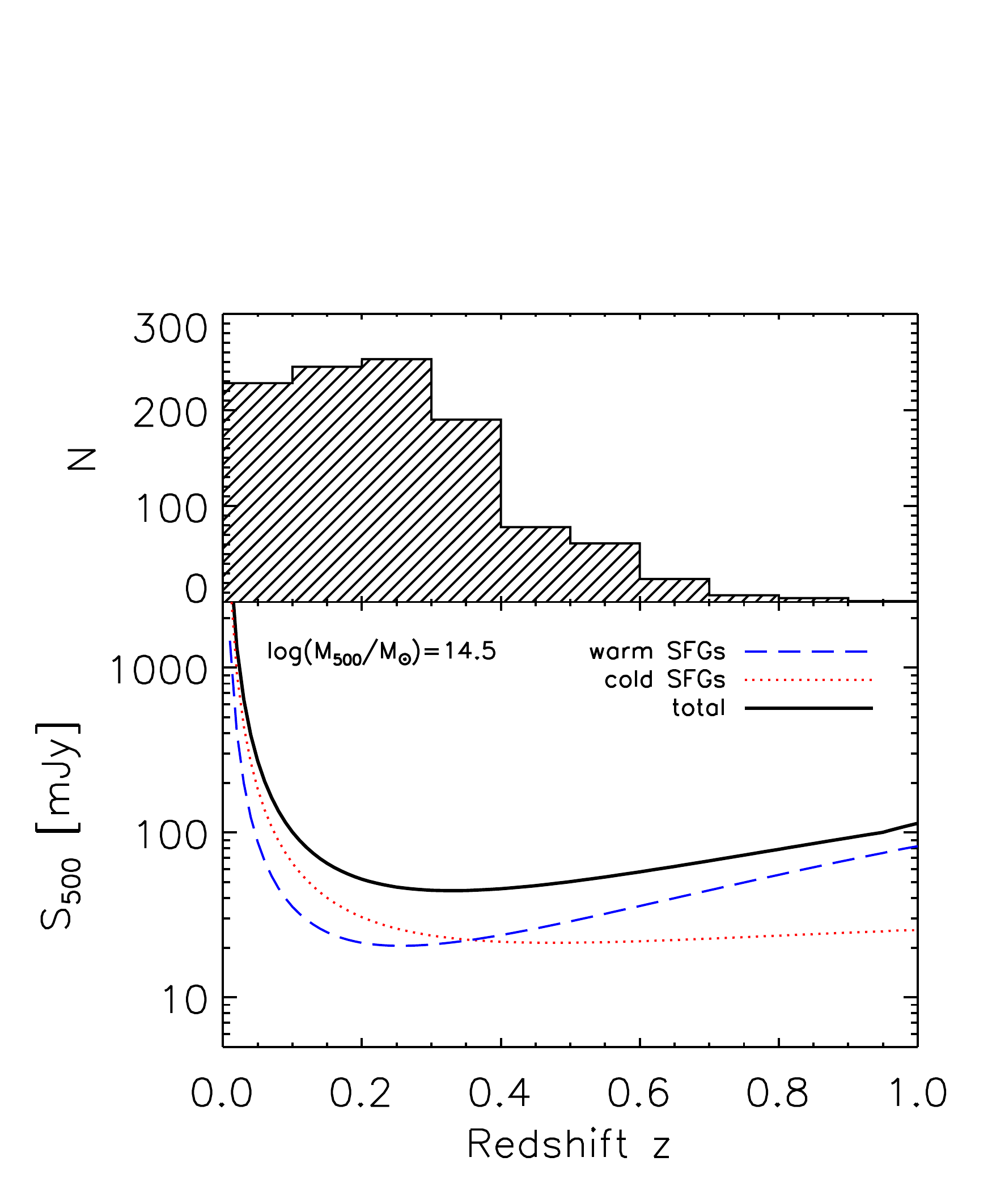}
\caption{{\it Top:} Redshift distribution of our PSZ2 sample (1091 clusters). {\it Bottom:} Predicted cluster dust flux density versus redshift from the~\cite{dezotti2016} model. The dust emission is integrated within a sphere of radius $R_{500}$ for a cluster of mass $M_{500}=10^{14.5} M_\sun$ in the 857~GHz \Planck\ band. We fixed $r_L=1$ (see Eq.~\ref{eq:lirlowz}) in this figure.}
              \label{fig:cai_flux_vs_nu}
\end{figure}

%--------------------------------------------------------------------
\section{Data}
\label{sec:data}

We used the all-sky maps of the \Planck\ High Frequency Instrument (HFI).  From August 2009 to January 2012, HFI observed the sky in six frequency bands centered on 100, 143, 217, 353, 545 and 857~GHz. We used the full mission temperature maps that can be downloaded from~\url{http://pla.esac.esa.int/pla/}. In our analysis, we have assumed that the beam for each map is Gaussian with full width at half maximum (FWHM) values of 9.659, 7.220, 4.900, 4.916, 4.675, 4.216~arcmin, respectively, for each band.

We also used the second \Planck\ catalog of SZ sources~\citep[PSZ2,][]{planckcatalogue2016}, keeping only sources with an assigned redshift (1093 objects), and re-extract their SZ signal using Multifrequency Matched Filters~\citep[MMF3, hereafter noted MMF for simplicity,][]{planckcatalogueesz,melin2012,planckcatalogue2014,planckcatalogue2016}. The re-extraction gives a positive signal-to-noise ratio (S/N) for 1091 objects, which constitutes the sample that we have adopted throughout this paper. Figure~\ref{fig:cai_flux_vs_nu} (top) shows the sample redshift distribution. 

The MMF re-extraction provides size-flux degeneracy curves for each source that we break using an independent X-ray size-flux relation \citep[combination of Eqs.~7 and 9 of][]{planckszcosmo2014}. This allows us to compute the mass proxy of each object, $M_{500}^{\rm Yz}$ (and the associated cluster size $\theta_{500}^{\rm Yz}$, since the redshift is known). More detail on the method of computing the mass proxy is given in Sec. 7.2.2 of~\cite{planckcatalogue2014}. We thus have 1091 objects with position, redshift $z$, mass $M_{500}^{\rm Yz}$ and size $\theta_{500}^{\rm Yz}$.\\

%--------------------------------------------------------------------
\section{Dust modeling}
\label{sec:dust_model}

Our baseline dust model is built on \Herschel\ observations of field and cluster galaxies~\citep{alberts2014,alberts2016}, and on the model from~\cite{cai2013} for the luminosity functions and spectral energy distributions. The model was developed for the prediction of cluster fluxes for the \CORE\ space mission~\citep{delabrouille2017, dezotti2016}. We briefly summarize its main elements.

The total comoving infrared field luminosity density, $\Psi_{\rm IR}(z)$, is computed using the model from~\cite{cai2013}. The model includes three populations of galaxies: the "warm" and "cold" populations dominate at $z<1$, the "spheroidal" population at $z>1.5$. Using the luminosity functions, $\Phi_i$, given by the model\footnote{http://staff.ustc.edu.cn/\textasciitilde{}zcai/galaxy\_agn/index.html}, we compute, for each redshift, the comoving infrared luminosity density, $\Psi_ i(z)$, contributed by each galaxy population~$i$:
\begin{equation}
 \Psi_{i}(z)={\int \Phi_{i}(\log L,z) \, L \, d\log L}.  
\label{eq:psiiz}
\end{equation}
for ${i}={\rm cold},{\rm warm},{\rm spheroidal}$. $\log$ is the base-10 logarithm. We then compute the total comoving infrared field luminosity density
\begin{equation}
 \Psi_{\rm IR}(z)=\sum_{i} \Psi_{i}(z).
\label{eq:psiirz}
\end{equation}

For $z<1.2$, \cite{alberts2014} measured the ratio of the mean infrared luminosity in clusters to that of galaxies in the field (see their Table~2)
\begin{equation}
f(z) = {L_{\rm cluster} \over L_{\rm field}} = {1540 \over 267} e^{-(0.76-0.42)t_{\rm Gyr}(z)} = 5.77 e^{-0.34t_{\rm Gyr}(z)}
\end{equation}
where $t_{\rm Gyr}(z)$ is the cosmic time in Gyr.

From $\Psi_{\rm IR}(z)$ and $f(z)$, we infer that the total infrared luminosity in the sphere of radius $R_{\Delta}$ for a cluster located at \mbox{$z<1.2$} is approximately
\begin{eqnarray}
L_{\Delta,{\rm tot},z<1.2} \approx \Psi_{\rm IR}(z) \times f(z) \times {\rho_{\rm cluster} \over \rho_{\rm mean}} \times V_{\rm comoving}, \nonumber \\
L_{\Delta,{\rm tot},z<1.2} \approx  {10^{14} M_\odot \over \rho_{\rm mean}(z=0)} \times \Psi_{\rm IR}(z)  \times f(z) \times {M_{\Delta} \over 10^{14} M_\odot},
\end{eqnarray}
where $V_{\rm comoving}=(1+z)^3 \times {4 \pi \over 3} R_{\Delta}^3$ is the comoving volume enclosed in the sphere of radius $R_{\Delta}$, $\rho_{\rm cluster}$ the cluster density, $\rho_{\rm mean}$ the matter density of the Universe at redshift $z$, $\Delta$ the overdensity with respect to critical density of the Universe at redshift $z$, and $M_{\Delta}$ the mass enclosed in the sphere of radius $R_{\Delta}$.

Fixing $\Delta=500$, we write
\begin{equation}
L_{500,{\rm tot},z<1.2} = r_L \times {10^{14} M_\odot \over \rho_{\rm mean}(z=0)} \times \Psi_{\rm IR}(z)  \times f(z) \times {M_{500} \over 10^{14} M_\odot},
\label{eq:lirlowz}
\end{equation}
with $r_L$ a normalization factor that we will determine from the 857~GHz flux of \Planck\ clusters (see Sect.~\ref{sec:constraints}). It is expected to be close to unity if \Herschel\ and \Planck\ data are consistent and our model is valid.

For $z>1.2$, the factor $f(z)$ does not apply because star formation in clusters matches that of field galaxies~\citep{alberts2016}; thus the IR luminosity reads
\begin{equation}
L_{500,{\rm tot},z>1.2} = r_L \times {10^{14} M_\odot \over \rho_{\rm mean}(z=0)} \times \Psi_{\rm IR}(z) \times {M_{500} \over 10^{14} M_\odot}.
\label{eq:lirhighz}
\end{equation}

We then compute, for each redshift, the fraction ${\rm p}_{i}(z)$ of the total luminosity contributed by the galaxy population ${i}$ as
\begin{equation}
{\rm p}_{\rm i}(z)={\Psi_{\rm i}(z) \over \Psi_{\rm IR}(z)}.
\end{equation}
The luminosity from each population is then given by
\begin{equation}
L_{500,{i}}[\nu (1+z)] = {\rm p}_{i}(z) \times L_{500,{\rm tot}} \times {\rm SED}_{i}[\nu (1+z)],
\end{equation}
where $\nu$ is the observation frequency and ${\rm SED}_{i}$ is the spectral energy distribution of the population $i$ normalized such that $\int {\rm SED}_{i}[\nu (1+z)] d\nu=1$~\citep[see e.g., Fig.~4 of][]{cai2013}.  The dust flux density for population $i$ is thus 
\begin{equation}
S_{500,{i}}(\nu) = {(1+z) L_{500,{i}}[\nu (1+z)] \over 4 \pi D_L^2(z)},
\label{eq:fluxir}
\end{equation}
with $D_L(z)$ the luminosity distance.
This model is identical to the model adopted in Sect. 4 of~\cite{dezotti2016}, except that we have introduced the normalization factor $r_L$. Fixing $r_L=1$ makes our model strictly identical to~\cite{dezotti2016}.

Figure~\ref{fig:cai_flux_vs_nu} (bottom) shows the predicted dust flux density integrated over the \Planck\ 857~GHz bandwidth, $S_{500}$, as a function of redshift for a cluster of mass $M_{500}=10^{14.5} M_\odot$ (thick solid line). It is essentially only composed of the warm (dashed blue line) and cold (dotted red line) components for redshifts $z<1$.  At $z>0.25$, the flux density increases with $z$ because of the increase in luminosity with $z$. At $z<0.25$, the luminosity distance dominates the redshift evolution.

Throughout this paper, we will use this model only in the redshift range $0<z<1$, which is relevant to the \Planck\ cluster catalog. But the model also includes the spheroidal component, which dominates at $z>1.5$, and we intend to use it in future work to examine the impact of dust emission on next generation cosmic microwave background (CMB)  experiments.

Our model gives global quantities (infrared luminosity), but it does not give any information on the spatial distribution of the dust emission in clusters.
In Sect.~\ref{sec:constraints}, we use \Planck\ PSZ2 clusters to jointly constrain the model normalization, $r_L$ (see Eq.~\ref{eq:lirlowz}), and the emission profile.

\vspace{1cm} % To avoid Latex crashing with hyperref

\section{\Planck\ constraints on the normalization and spatial profile of cluster dust emission}
\label{sec:constraints}

%-------------------------------------- Two column figure (place early!)
\begin{figure*}
%\begin{figure}
\centering
\includegraphics[width=0.45\hsize]{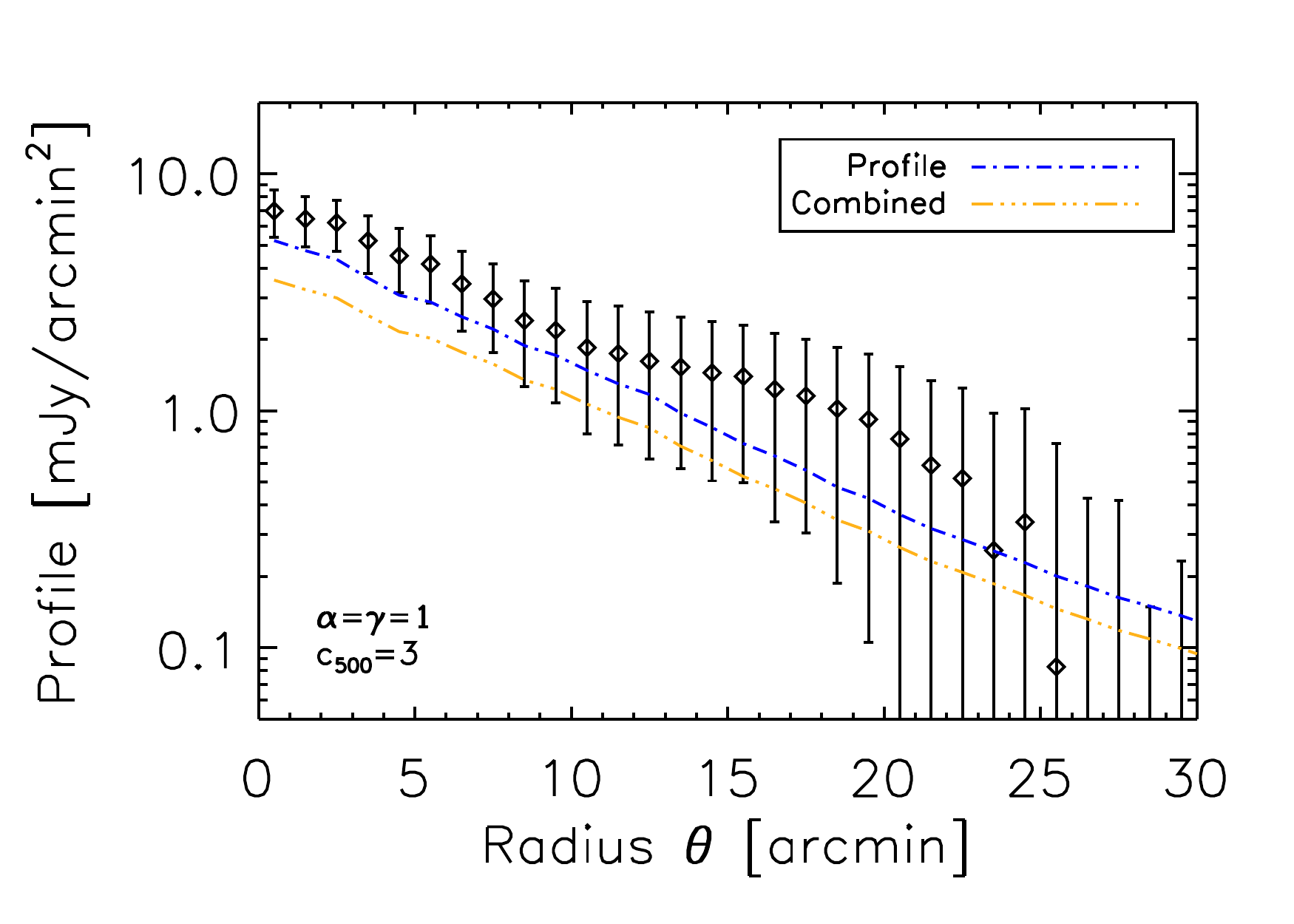} \includegraphics[width=0.45\hsize]{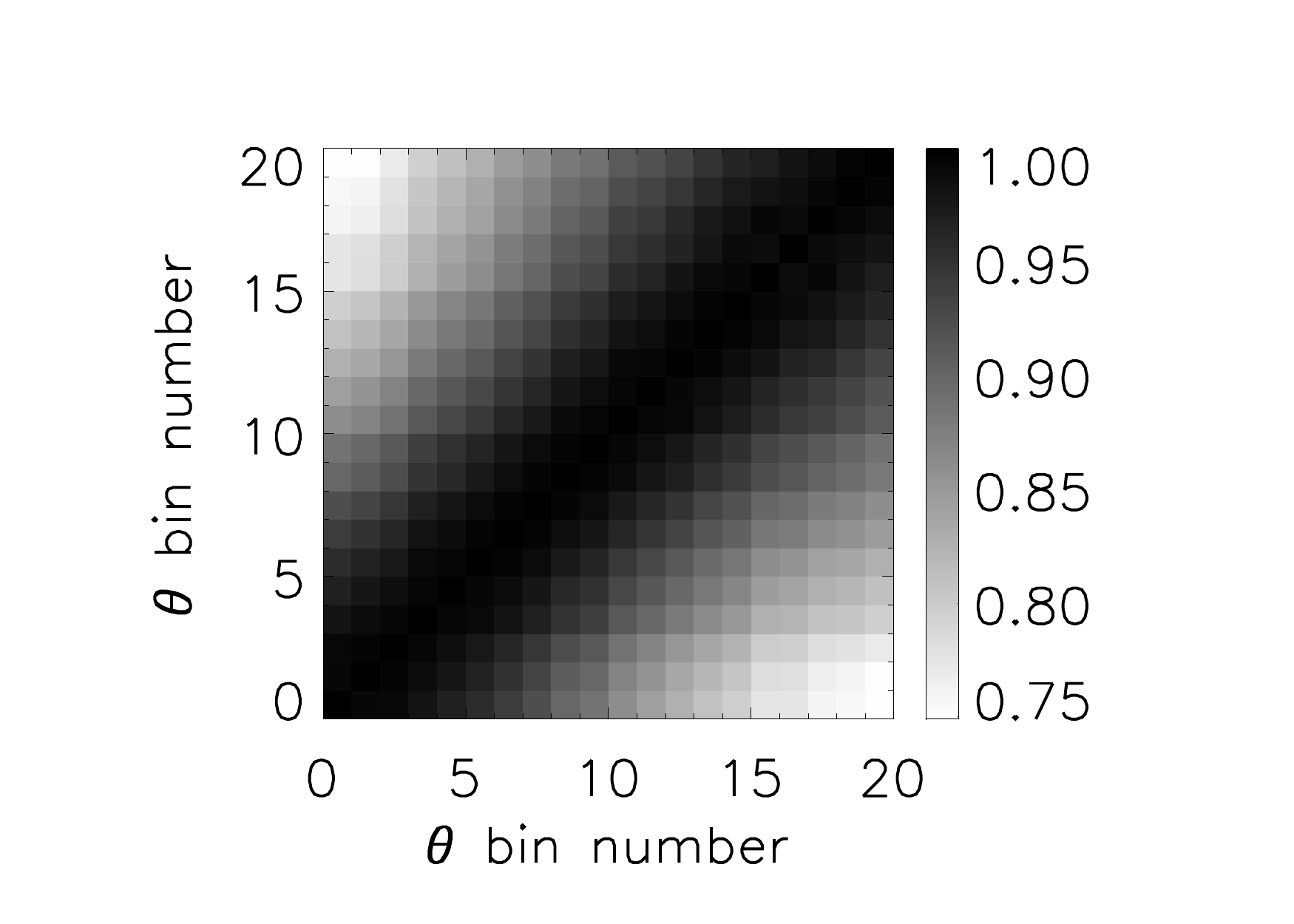}
\caption{{\it Left:} Stacked PSZ2 profiles in the 857~GHz \Planck\ band (black diamonds) and best fit profile (blue dash-dotted line). Error bars are determined from bootstrap realizations.  The data points are strongly correlated, as shown in the right panel. The orange dash-double dotted line shows the best fit profile obtained when adjusting jointly the stacked PSZ2 profile and the inverse-variance weighted average matched filter flux in the \Planck\ 857~GHz band. 
{\it Right:} Correlation matrix (diagonal normalized to unity) of the 20 first bins of the stacked profiles starting from the center (bin zero is the most central bin). The bins are strongly correlated ($>75\%$).}
              \label{fig:stacked_profile}
\end{figure*}
%\end{figure}
%

We describe the three dimensional dust emission profile with a Generalized Navarro-Frenk-White (GNFW) profile~\citep{nagai2007}:
\begin{equation}
\label{eq:gnfw}
P(r) \propto {1 \over (r / r_s )^\gamma \left [ 1 + ( r / r_s )^\alpha  \right ]^{\beta - \gamma \over \alpha}},
\end{equation}
where $\alpha$, $\beta$ and $\gamma$ are, respectively, the intermediate, external and central slopes, $r_s=R_{500}/c_{500}$ is the scale radius and $c_{500}$ is the concentration parameter.  Dark matter profiles of massive relaxed clusters follow a standard NFW~\citep{navarro1996} profile with $\alpha=\gamma=1$ and $\beta=3$. For nearby relaxed galaxy clusters, \cite{pointecouteau2005} have constrained $c_{200}=4.61 \pm 0.12$, which we convert to $c_{500}=3.03 \pm 0.08$.  

We used two observables, the stacked profile (Sect.~\ref{sec:stack}) and the inverse-variance weighted average matched filter flux (Sect.~\ref{sec:ivmff}), to constrain the normalization, $r_L$, in Eq.~(\ref{eq:lirlowz}) and the dust emission profile. The stacked profile does not provide enough information to simultaneously constrain all of the GNFW parameters. We therefore fixed $\alpha=\gamma=1$ and $c_{500}=3$, and leave only the external slope, $\beta$, free.  A value of $\beta$ larger (or smaller) than three indicates that the profile is steeper (or shallower) than the dark matter profile of massive relaxed clusters. The constraints from the stacked profile and the inverse-variance weighted matched filter can be first compared and then combined (Sect.~\ref{sec:combi}).

\subsection{Stacked profile}
\label{sec:stack}

Our first observable is the stacked profile of the PSZ2 clusters (see Sect.~\ref{sec:data}), which we constructed following the same methodology as~\cite{hurier2016}. For each cluster, we computed the unweighted mean flux of the 857~GHz map in annuli of width $\Delta \theta=1 \, {\rm arcmin}$, starting from $\theta=0$ (the SZ center) to $\theta=30 \, {\rm arcmin}$. We removed the offset using the mean value of the pixels between $\theta=30 \, {\rm arcmin}$ and 60~arcmin. We then took the mean of all the profiles in each annulus. 

This stacked profile is shown as the black diamonds in the lefthand panel of Fig.~\ref{fig:stacked_profile}. The error bars were obtained as the standard deviation of 10,000 bootstrap realizations. Stacking the profile in angular radii $\theta$ mixes different physical scales. This procedure thus introduces correlations between the bins, which can be estimated from the bootstraps. The righthand panel of Fig.~\ref{fig:stacked_profile} shows the correlation matrix, and we see that the data points are indeed strongly correlated ($>75\%$). 

\begin{figure}
\centering
\includegraphics[width=\hsize]{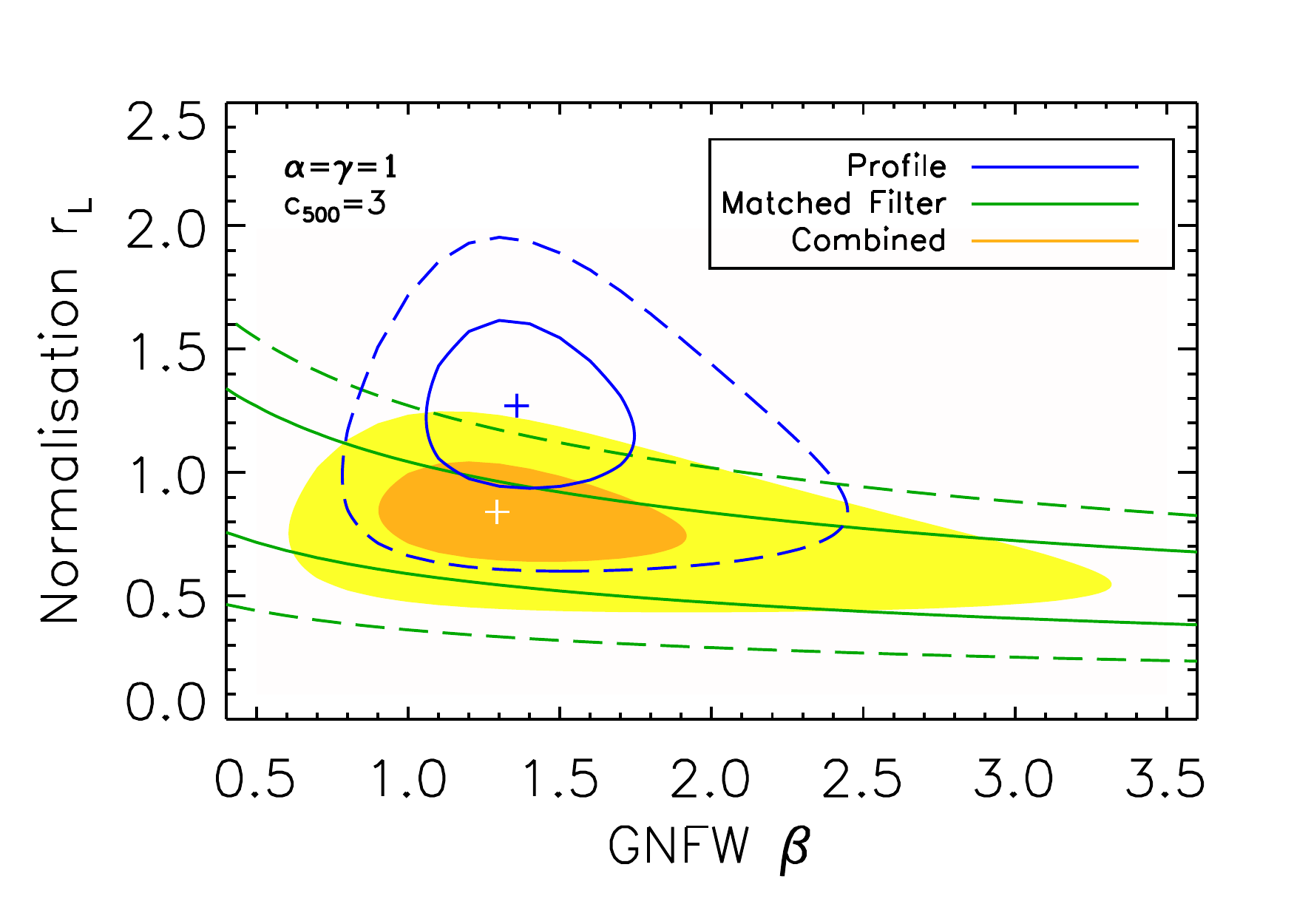}
\caption{Contours at 68\% and 95\% C.L. on the normalization, $r_L$, of the infrared $L_{500,{\rm tot}}-M_{500}$ relation and on the external slope, $\beta$, of the spatial profile of the dust emission. Constraints are obtained from the stacked profile (blue) and from the inverse-variance weighted average matched filter flux (green), both in the \Planck\ 857~GHz band. The combined constraint is shown as filled orange and yellow contours. The blue and white crosses shows the best value for the profile and combined fits, respectively.}
              \label{fig:rL_vs_beta}
\end{figure}

We fit the observed profile using a stacked GNFW profile. For each cluster, a GNFW profile is scaled to $\theta_{500}^{\rm Yz}$ and normalized using Eq.~(\ref{eq:fluxir}). We then applied the same averaging procedure as for the data. We fixed all the parameters and let only $r_{\rm L}$ and $\beta$ vary. The 68\%/95\% confidence limits (C.L.) on $r_{\rm L}$ and $\beta$ are shown as the solid and dashed blue lines respectively in Fig.~\ref{fig:rL_vs_beta}. The best fit values are given in the first row of Table~\ref{tab:bestfitvalues}.  The normalization, $r_{\rm L}=1.27 \substack{+0.34 \\ -0.33}$, is compatible with one, the value determined from \Herschel\ data.  The slope parameter, $\beta = 1.36 \substack{+0.39 \\ -0.30}$, is significantly lower than three, indicating that the dust profile is shallower than the matter profile. The fit of the stacked profile is shown as the blue dash-dotted line in the lefthand panel of Fig.~\ref{fig:stacked_profile}. We note that it is systematically below the majority of the data points. This is due to the strong correlation between the points, as given by the correlation matrix in the righthand panel of the figure.

\begin{table*}
%\begin{table}
\centering
\begin{tabular}{| c | c | c | c | c | c |} 
\hline
\rule[-0.2cm]{0cm}{0.6cm} & $r_L$ & $\beta$ & $\chi^2$ & \# $d.o.f.$ & $\chi^2/d.o.f.$ \\
\hline
\rule[-0.2cm]{0cm}{0.6cm} Profile & $1.27 \substack{+0.34 \\ -0.33}$ & $1.36 \substack{+0.39 \\ -0.30}$ & 20.65 & 18 & 1.15 \\
\hline
\rule[-0.2cm]{0cm}{0.6cm} Combined & $0.84 \substack{+0.20 \\ -0.20}$ & $1.29 \substack{+0.63 \\ -0.39}$ & 26.63 & 19 & 1.40 \\
\hline
\end{tabular}
\vspace{0.2cm}
\caption{Best fit values for the normalization, $r_L$, of the infrared $L_{500,{\rm tot}}-M_{500}$ relation and for the external slope, $\beta$, of the spatial profile of the dust emission. Errors are 68\% C.L.}             % title of Table
\label{tab:bestfitvalues}
\end{table*}
%\end{table}

\subsection{Inverse-variance weighted average matched filter flux}
\label{sec:ivmff}

Our second observable is the inverse-variance weighted average matched filter flux of the PSZ2 clusters measured in 857~GHz maps. We extract individual cluster flux and associated error in the 857~GHz map using a single frequency matched filter~\citep{melin2006}. We fix the position to the SZ center and the size to $\theta_{500}^{\rm Yz}$, and we adopt the universal pressure profile from~\cite{arnaud2010}. We also perform the flux extraction at the five other \Planck\ HFI frequencies, although we do not use them to constrain $r_{\rm L}$ and $\beta$. Results are shown in Fig.~\ref{fig:mf_flux_vs_nu} as black diamonds. The error bars are obtained from the standard deviation of 10,000 bootstrap realizations of the inverse-weighted average.

We then fit the GNFW model to the 857~GHz data point by applying the matched filter to the model and averaging as done on the real data.  The constraints are shown as green contours in Fig.~\ref{fig:rL_vs_beta}. There is no absolute minimum, since for each value of $\beta$ we can find $r_{\rm L}$ which adjusts the average matched filter flux at 857~GHz. The contour is thus a valley.

\subsection{Combination}
\label{sec:combi}

The $r_{\rm L}$ values preferred by the inverse-variance weighted average matched filter flux are lower than the values preferred by the stacked profile, but the two observations are nevertheless compatible.
We used the 10,000 bootstrap realizations to estimate the correlation of the two observables and then combine them into a signal constraint. The resulting 68\% and 95\% C.L. are shown as the orange and yellow filled contours in Fig.~\ref{fig:rL_vs_beta}. The corresponding best fit is marked with a white cross and is given in the second row of Table~\ref{tab:bestfitvalues}. The result, $r_{\rm L}=0.84 \pm 0.20$ is compatible with one, as for the stacked profile constraint. The value for $\beta$ is fully driven by the stacked profile because the average matched filter flux does not constrain it. 

The best fit for the stacked profile has 18 degrees of freedom (d.o.f). corresponding to 20 radial bins minus two parameters. The best fit for the combined constraint is 19 d.o.f. corresponding to 20 radial bins plus 1 bin averaged matched filter flux minus two parameters. The reduced $\chi^2$ is acceptable ($\chi^2/d.o.f.$=1.15) for the profile only fit and shows some small tension between the two measurements for the combined case ($\chi^2/d.o.f.$=1.40).

\begin{figure}
\centering
\includegraphics[width=\hsize]{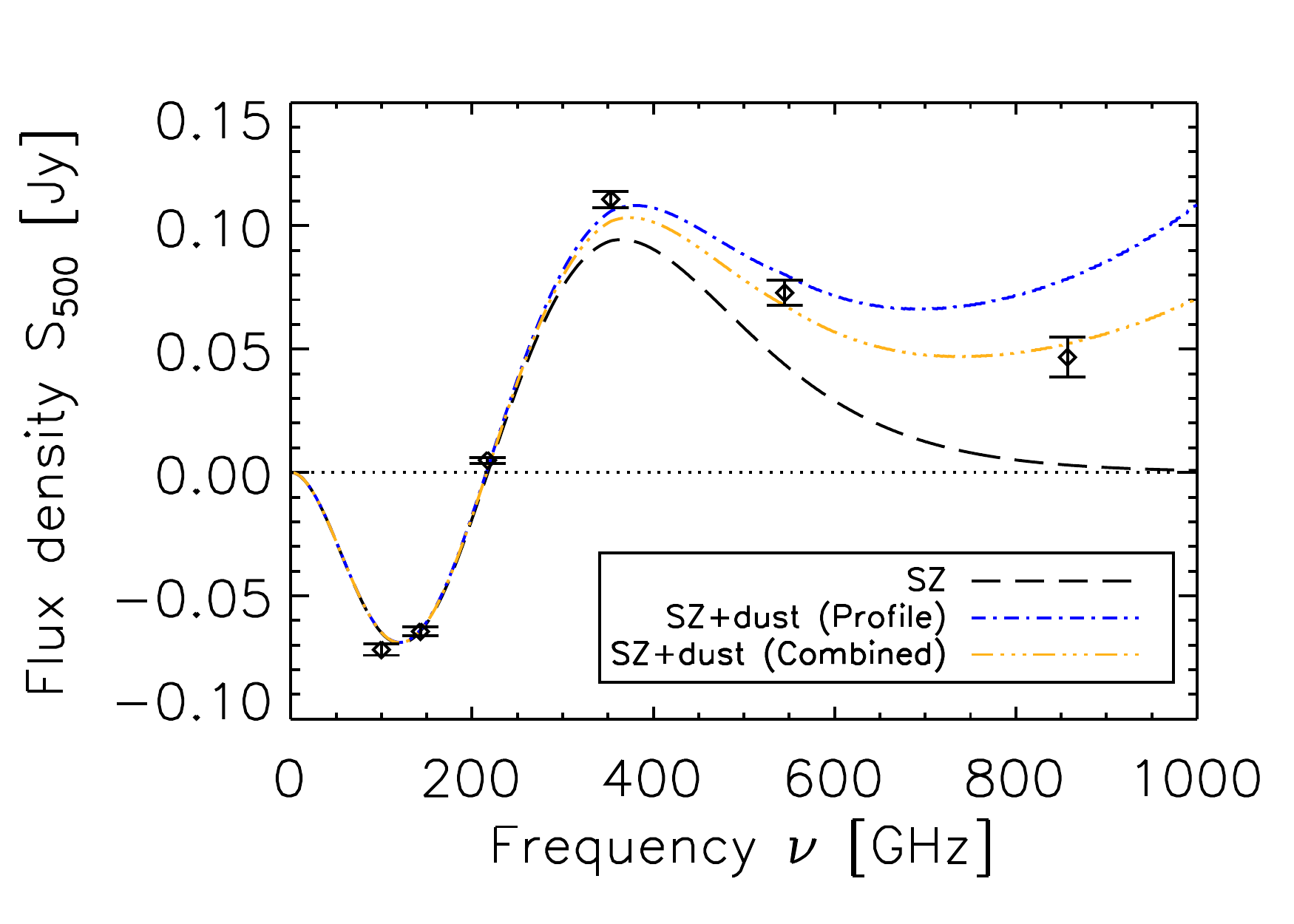}
\caption{Inverse-variance weighted matched filter flux in the HFI maps (black diamonds) and associated bootstrap errors. The profile used in the matched filter is the universal pressure profile from~\cite{arnaud2010}. The black dashed line shows the SZ contribution calculated by inverse-variance averaging the \Planck\ Compton $y$ values. Blue dash-dotted and orange dash double-dotted lines show the SZ+dust models (blue: dust best fit from stacked PSZ2 profiles in the \Planck\ 857~GHz band, orange: dust combined best fit from stacked PSZ2 profile and inverse-variance weighted matched filter flux in the \Planck\ 857~GHz band).}
              \label{fig:mf_flux_vs_nu}
\end{figure}

The dust profile corresponding to the combined best fit is shown as the orange dash-double dotted line in Fig.~\ref{fig:stacked_profile}. The inverse-variance weighted matched filter flux for the profile (combined best fit) model is shown as the blue dash-dotted (orange dash double-dotted) line for SZ+dust in Fig.~\ref{fig:mf_flux_vs_nu}, to be compared to the SZ-only signal shown as the black dashed curve. The blue line is significantly ($3.9\sigma$) higher than the measurement in the 857~GHz band. In the 857~GHz band, the orange line is in good agreement with the data by construction.

Although the combined fit is performed using 857~GHz data only, the agreement at lower frequencies is good. This demonstrates that the galaxy populations of the~\cite{dezotti2016} model and their spectral energy distributions provide a satisfactory description of the frequency dependance of the dust emission of \Planck\ clusters. The comparison between the spectral energy distributions of galaxies in the~\cite{dezotti2016} model and the spectral energy distribution of \Planck\ clusters determined by~\cite{comis2016} is discussed in Appendix~\ref{app:comis}.\\

We adopt the result of the combined fit (second row of Table~\ref{tab:bestfitvalues}) as our fiducial dust model. The corresponding model parameters are $r_{\rm L}=0.84$ in Eq.~(\ref{eq:lirlowz}) and $\beta=1.29$ in Eq.~(\ref{eq:gnfw}) (with $\alpha=\gamma=1$, $c_{500}=3$ fixed).

%--------------------------------------------------------------------
\section{Impact on \Planck\ cosmological results}
\label{sec:impact}

\Planck\ cosmological analyses with clusters (in particular, cluster extraction and cosmological constraints from cluster counts) do not take into account dust emission in clusters.
The omission of this emission may possibly impact the cluster physical parameter recovery (cluster size and flux) and the survey completeness. In Sect.~\ref{sec:clusterfluxsize}, we use our fiducial dust model built in Sect.~\ref{sec:constraints} to study the effect of cluster dust emission on size and flux recovery. In Sect.~\ref{sec:completeness}, we calculate the effect of cluster dust emission on the \Planck\ completeness and show the impact on cosmological parameter determination.

\subsection{Cluster size and flux recovery}
\label{sec:clusterfluxsize}

The \Planck\ beams (FWHM ranging from 9.6 to 4.2 arcmin between 100 and 857~GHz) are larger than the typical cluster extent (1~arcmin), meaning that \Planck\ provides weak constraints on cluster size. As a direct consequence, blind fluxes are only weakly constrained by the extraction tools. This problem is often referred as the "size-flux degeneracy"~\citep[see e.g., introduction of Sect. 7.2 of][]{planckcatalogue2014}. 

The \Planck\ collaboration noticed that blindly recovered cluster sizes are over estimated on average with respect to the sizes estimated from X-ray observations. This size over-estimation translates into an over-estimation in the blind flux relative to expectations based on the X-rays. For this reason, the \Planck\ collaboration computed cluster flux fixing the size from X-ray measurements~\citep[Sect. 7.2.1 of][]{planckcatalogue2014} or adopting a size-flux relation from X-ray as a prior to break the \Planck\ size-flux degeneracy. The latter approach is used to derive the "mass proxy" $M_{500}^{\rm Yz}$~\citep[Sect. 7.2.2 of][]{planckcatalogue2014} that we use in this paper. However, this size overestimation is not present in millimeter simulations which include SZ as the only cluster emission and for which the simulated SZ profile perfectly matches the profile assumed in the extraction tool~\citep[Fig. 8 of][]{melin2006}.
In this section, we examine the over-estimation seen in the \Planck\ data, looking to see if it is related to some additional cluster component, such as dust, or is linked to the profile assumed for the extraction.

We simulate 1091 clusters with the same masses and redshifts as the PSZ2. We inject them randomly into \Planck\ frequency maps, outside the 85\% survey mask\footnote{Used in the construction of the PSZ2 cluster catalog~\citep{planckcatalogue2016}.} to avoid contamination by Galactic dust and outside a PSZ2 cluster mask\footnote{Rejects circular regions of size $3\theta_{500}^{\rm Yz}$ around the PSZ2 clusters.} to avoid contamination by real clusters. We model the SZ emission using the universal pressure profile~\citep[UPP,][]{arnaud2010} or the \Planck\ pressure profile~\citep[PlanckPP,][]{planckprof2013}. Although the PlanckPP is consistent with the UPP in the inner cluster regions ($R<R_{500}$), it is significantly more extended to larger radii~\citep[$R_{500}<R<3R_{500}$, see Fig. 4 left of][]{planckprof2013}. Thus, assuming the UPP for extracting clusters well described with a PlanckPP could possibly lead to a size over-estimation. 

We then modeled the dust emission using our combined best fit described in Sect.~\ref{sec:dust_model}. We simulate the four possible combinations (UPP and PlanckPP, with and without dust) and extract cluster size and flux using the MMF. For the cluster size, $\theta_{500}$, we use a grid of 32 filter sizes equally spaced on a logarithmic scale and ranging from 0.94 to 35.31~arcmin. We searched for the cluster position as a maximum of S/N in a circle of radius 20~arcmin around the real or injected position and adopt the UPP in the MMF for the four cases. We then compared the recovered size and flux to their input values to see if we could identify the origin of the blind size and flux over-estimation found in the data.

\begin{figure*}
%\begin{figure}
\centering
\includegraphics[width=0.45\hsize]{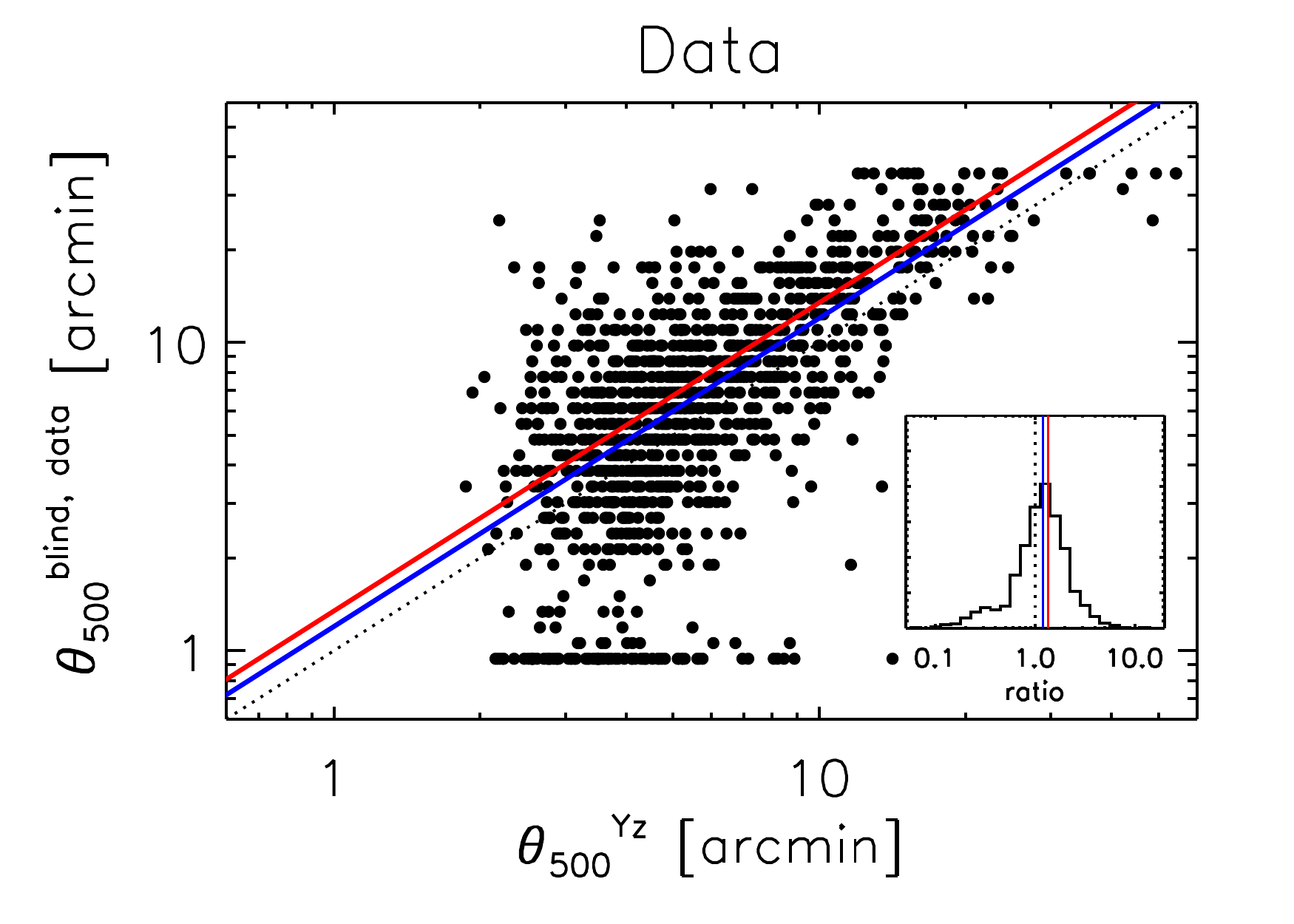} \includegraphics[width=0.45\hsize]{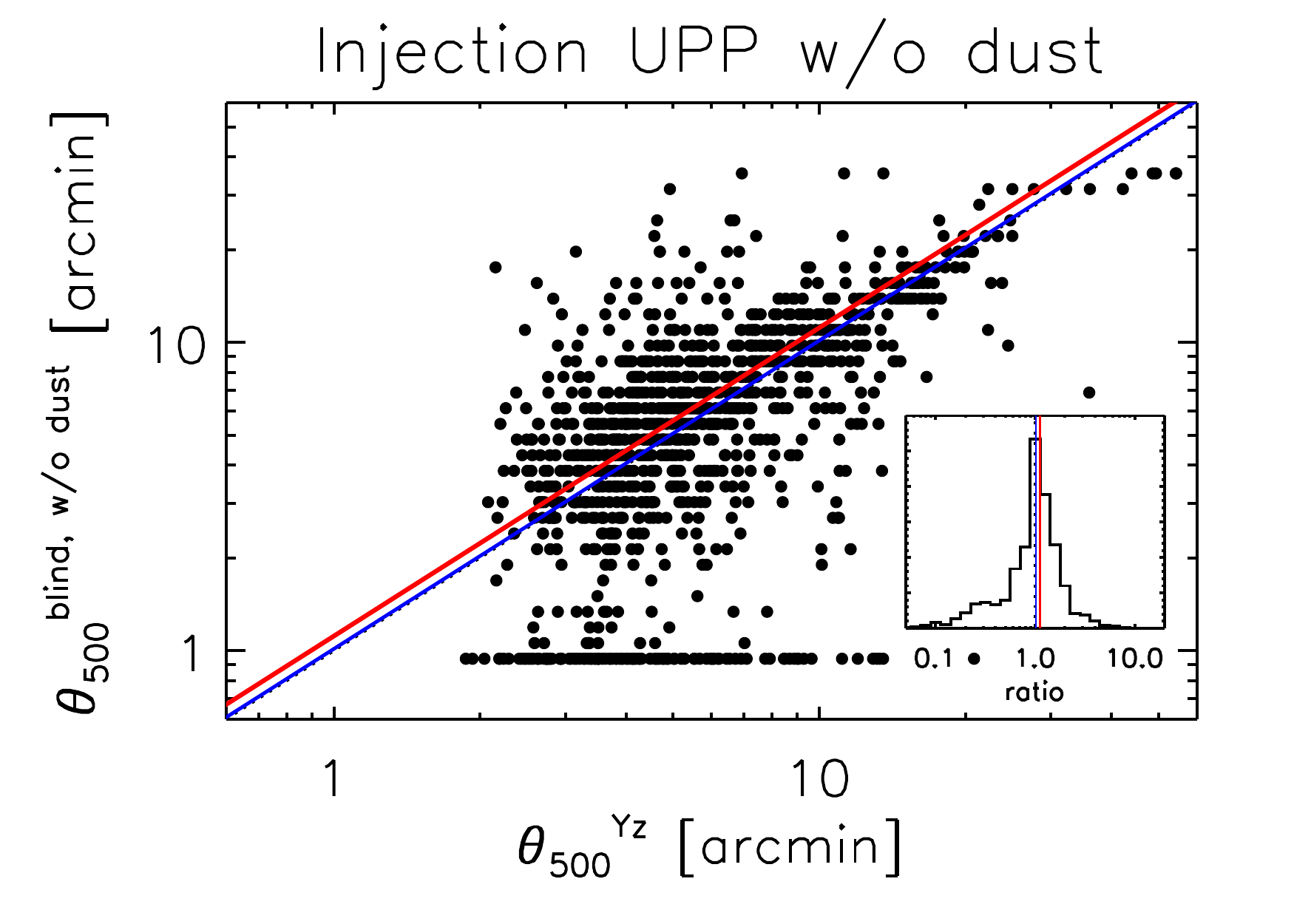}
\includegraphics[width=0.45\hsize]{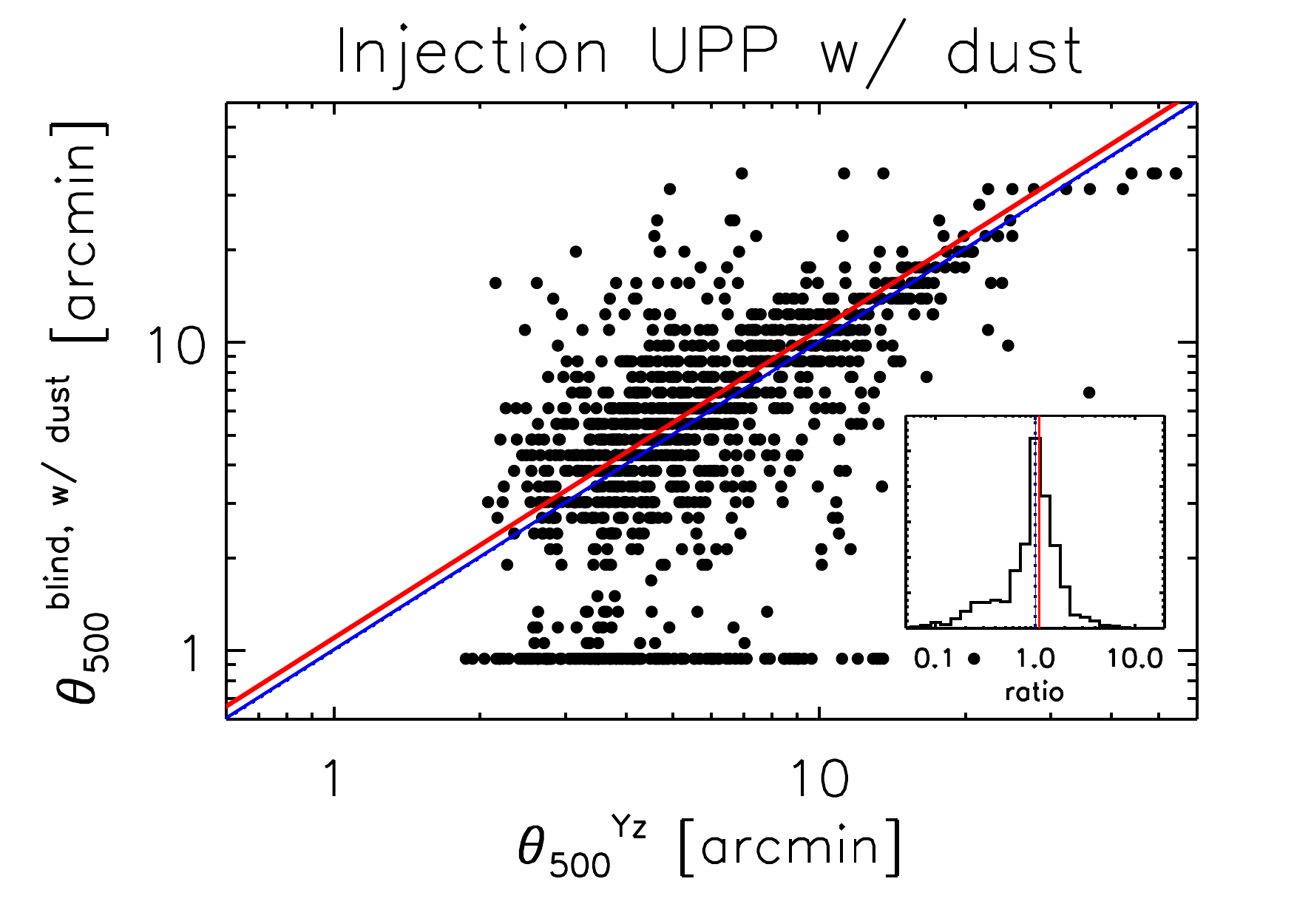} \includegraphics[width=0.45\hsize]{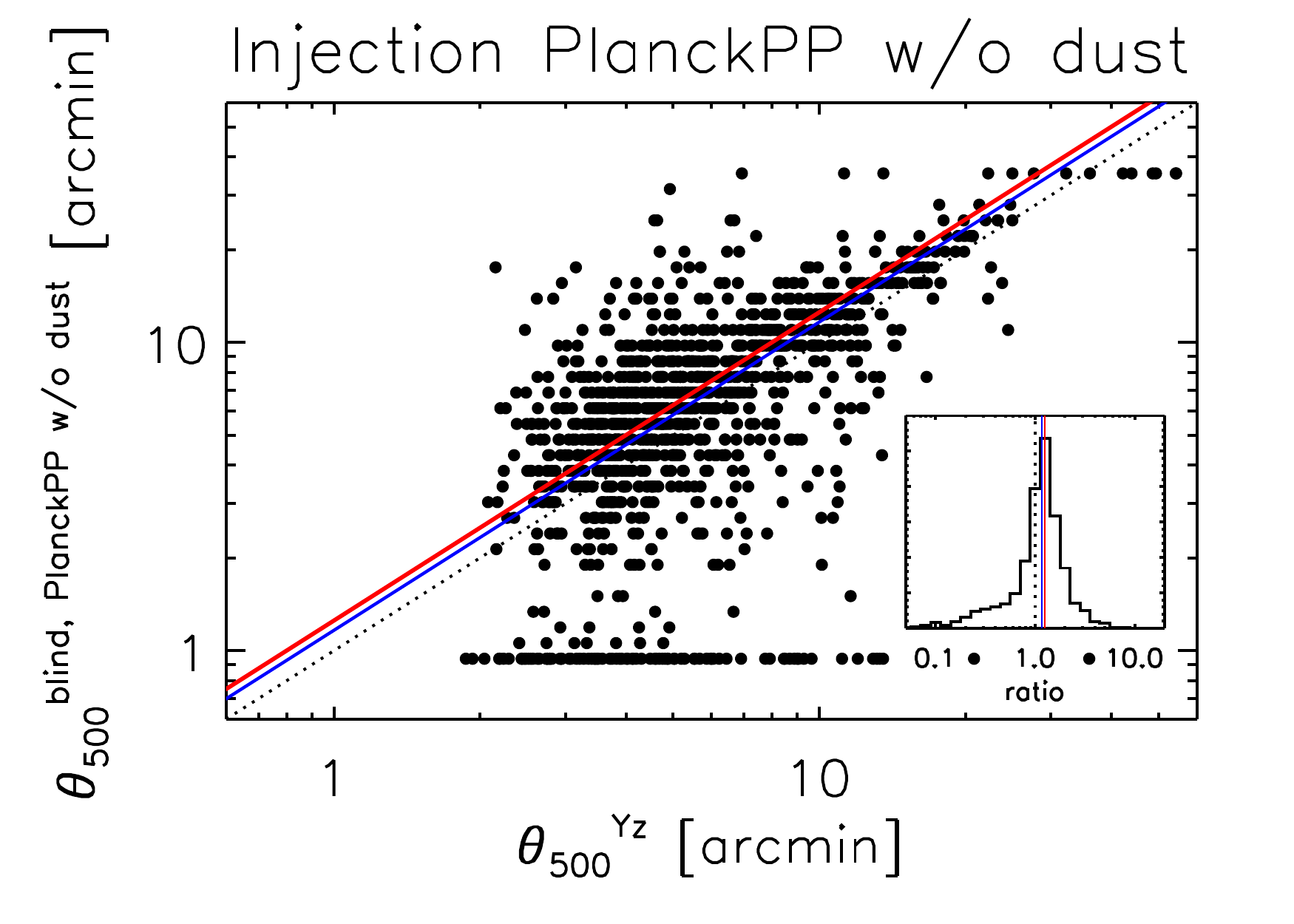}
\caption{Cluster size $\theta_{500}^{\rm blind}$ extracted blindly versus $\theta_{500}^{\rm Yz}$ from the \Planck\ mass proxy. The universal pressure profile is used in the matched filter for the extraction in the four panels. {\it Top left:} Extraction from \Planck\ data. The blind sizes are systematically overestimated with respect to the size derived from the mass proxy (equivalent to a \XMM\ size). The mean (median) of the ratio $\theta_{500}^{\rm blind} \over \theta_{500}^{\rm Yz}$ is 1.35 (1.20) and is displayed as the red (blue) line. The thickness of the line encapsulates the 68\% error on the mean (median) calculated with bootstrap. The histogram of this ratio is shown in the inset. {\it Top right:} Extraction from injections in \Planck\ data assuming that the SZ emission follows the universal pressure profile~\citep[UPP,][]{arnaud2010}. No dust emission was included. The value for the mean (median) is 1.12 (1.02). There is no strong overestimation of the size as on the actual data. {\it Bottom left:} Same as top right but adding the dust component based on the best combined fit (white cross in Fig.~\ref{fig:rL_vs_beta}). The mean (median) is 1.10 (1.01). The impact of the dust component on the size estimation is negligible. {\it Bottom right:} Same as top right but using the \Planck\ pressure profile instead of the UPP to simulate clusters. The mean (median) is 1.25 (1.16). The blind sizes are overestimated as for the \Planck\ data, although the histogram in the inset is less dispersed. The dotted line in all four panels is the equality line.}
              \label{fig:recovered_sizes}
%\end{figure}              
\end{figure*}

Results are shown in Fig.~\ref{fig:recovered_sizes} for the size. The top left panel shows the extraction at the location of the actual PSZ2 clusters. The recovered size $\theta_{500}^{\rm blind,data}$ is weakly constrained and is significantly biased high with respect to the size derived from the mass proxy $\theta_{500}^{\rm Yz}$. The mean (median) of the ratio of the two quantities is 1.35 (1.20). The thickness of the line encapsulates the 68\% error on the mean (median) calculated with bootstrap. The inset shows the histogram of the distribution, which peaks above one. The recovered size $\theta_{500}^{\rm blind,data}$ is discretized and corresponds to the values adopted for our grid. One can also notice that the algorithm sometimes fails to recover cluster size and falls onto the grid limits. 

The top right panel shows the extraction at the location of the injections with the UPP and without dust emission. As already noticed in~\cite{melin2006}, although the recovered size  $\theta_{500}^{\rm blind, \, w/o \, dust}$ is weakly constrained, it is significantly less biased with respect to the injected size $\theta_{500}^{\rm Yz}$. The mean (median) of the ratio is 1.12 (1.02) and the histogram peaks around unity. Again, the thickness of the line encapsulates the 68\% error on the mean (median) calculated with bootstrap. We tested this bootstrap error on the mean (median) by performing ten injections of the PSZ2 clusters and in computing the standard deviation of the mean (median) values across the ten corresponding extractions. The standard deviation across the ten extractions is in very good agreement with the bootstrap error on a single extraction, with a value of 0.02 for both the mean and the median.

The result of adding dust emission to SZ emission in simulated clusters is shown in the bottom left panel for the UPP. The result is essentially identical to the UPP without dust. The mean (median) of the ratio is 1.10 (1.01) and the histogram peaks around unity. 

The effect of changing the SZ profile to the PlanckPP is shown in the bottom right panel, without dust emission. The recovered size is overestimated with a mean (median) ratio equal to 1.25 (1.16), close to the value observed in the actual data. The result of including dust with the PlanckPP is not shown: it is almost identical to the PlanckPP without dust as for the UPP.

This test demonstrates that dust emission has no impact on cluster size estimate with the MMF. It also indicates that the size overestimation may find its origin in the profile mismatch between actual clusters and the UPP. Indeed, adopting the PlanckPP in the simulations and extracting clusters using the UPP reproduces the bias observed in the data. The dispersion of the histogram in the bottom right panel (PlanckPP without dust) is smaller than that of the histogram in the top left panel (actual data). This could be due to the dispersion in the actual pressure profiles which is not included in the simulations, the PlanckPP being the average value.

The size overestimation shown in Fig.~\ref{fig:recovered_sizes} directly impacts the flux. Thus the flux estimation depends on the profile assumed for the injected model. The results are shown in Appendix~\ref{app:psz2_flux} (Fig.~\ref{fig:recovered_fluxes}). As for the size, the flux is overestimated for the actual PSZ2 and the PlanckPP case (without and with dust), but not overestimated for the UPP (without and with dust).

\begin{figure}
\centering
\includegraphics[width=\hsize]{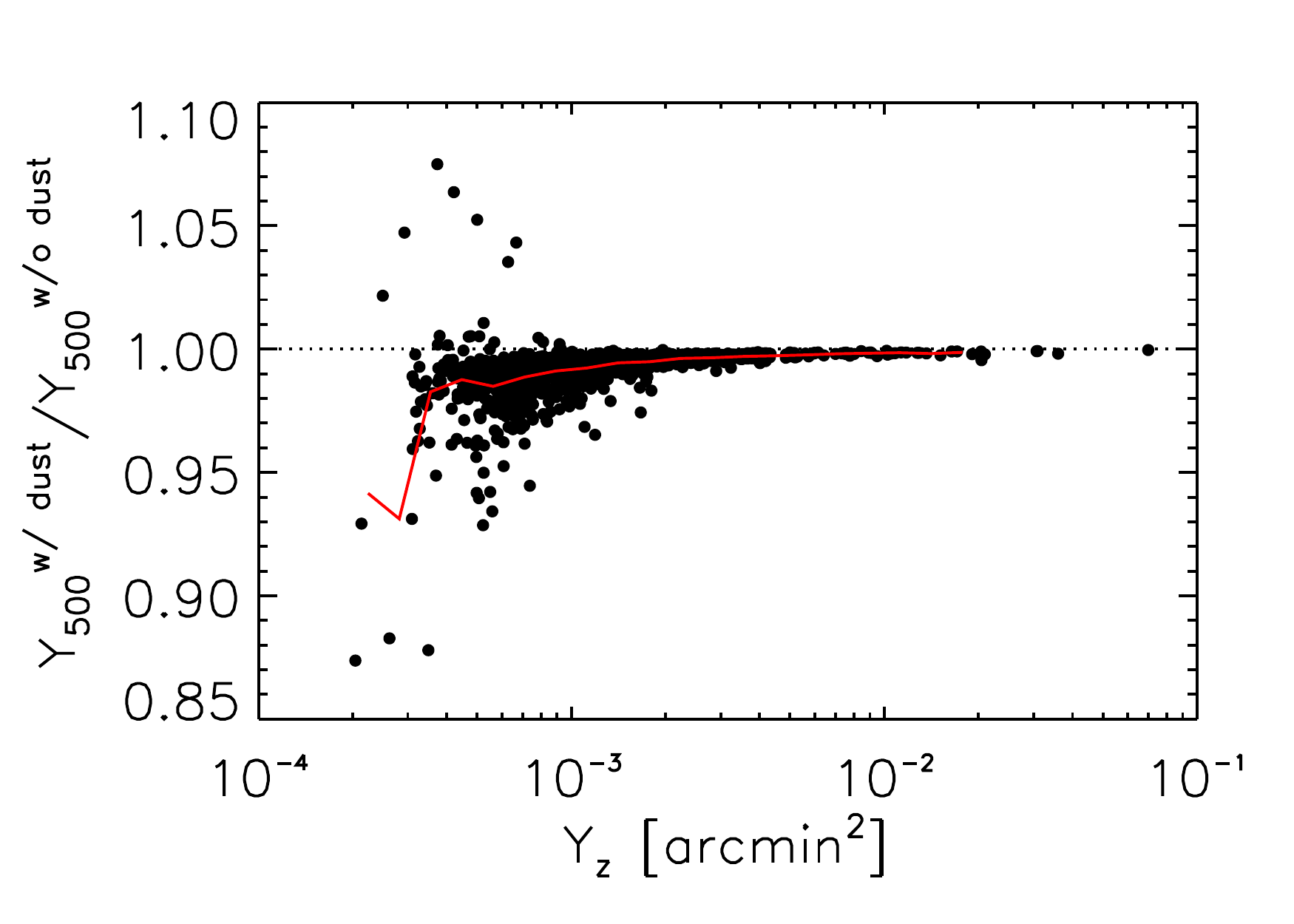}
\caption{Ratio of extracted MMF fluxes, $Y_{500}^{\rm w/ \, dust}/Y_{500}^{\rm w/o \, dust}$, when fixing position and size for clusters simulated with and without dust as a function of injected SZ flux $Y_z$. The red line is the raw mean value. The impact of dust emission is negligible for bright clusters ($Y_z>10^{-3} \, {\rm arcmin}^2$) and increases to $\sim 2\%$ with decreasing flux down to $Y_z=5 \times 10^{-4} \, {\rm arcmin}^2$.}
\label{fig:dust_flux_fixedposandsize}
\end{figure}

Finally, we examined the impact of dust emission on the PSZ2 flux estimation when fixing both cluster position and size. Results are shown in Fig.~\ref{fig:dust_flux_fixedposandsize}. The impact of dust emission on flux estimation is negligible ($<1\%$) for bright clusters ($Y_z>10^{-3} \, {\rm arcmin}^2$). The bias due to dust then increases from $<1\%$ to $\sim 2\%$ with decreasing flux  from $Y_z=10^{-3} \, {\rm arcmin}^2$ to $5 \times 10^{-4} \, {\rm arcmin}^2$.

\subsection{\Planck\ cluster completeness and cosmological constraints}
\label{sec:completeness}

We now investigate the impact of cluster dust emission on the \Planck\ cluster catalog completeness, and then on the measurement of cosmological parameters from cluster counts.

We randomly drew cluster redshifts and masses from the~\cite{tinker2010} mass function, model SZ emission with the UPP and normalize the flux using the $Y-M$ relation from~\cite{arnaud2010}. We adjusted the mass bias to $1-b=0.65$~\citep[see Eq.~7 of][]{planckszcosmo2014} to match the model counts in our adopted cosmology to the observed counts. We injected the clusters into the \Planck\ frequency maps at random locations outside the 65\% cosmological mask\footnote{Used in the Planck cluster cosmology analysis~\citep{planckszcosmo2016}.} and outside the same PSZ2 cluster mask as in Sect.~\ref{sec:clusterfluxsize} to avoid contamination by real clusters.  We considered two cases: with and without inclusion of dust emission in addition to the SZ signal. We then used the MMF algorithm to extract clusters blindly, following the same procedure as for the \Planck\ analyses~\citep[][]{planckcatalogueesz,planckcatalogue2014,planckcatalogue2016}. We perform ten such injections for each case. In order to improve the statistics at high redshift, we also simulated, injected and extracted ten additional independent catalogs containing only clusters at $z>0.5$, but with ten times higher density.

\begin{figure}
\centering
\includegraphics[width=\hsize]{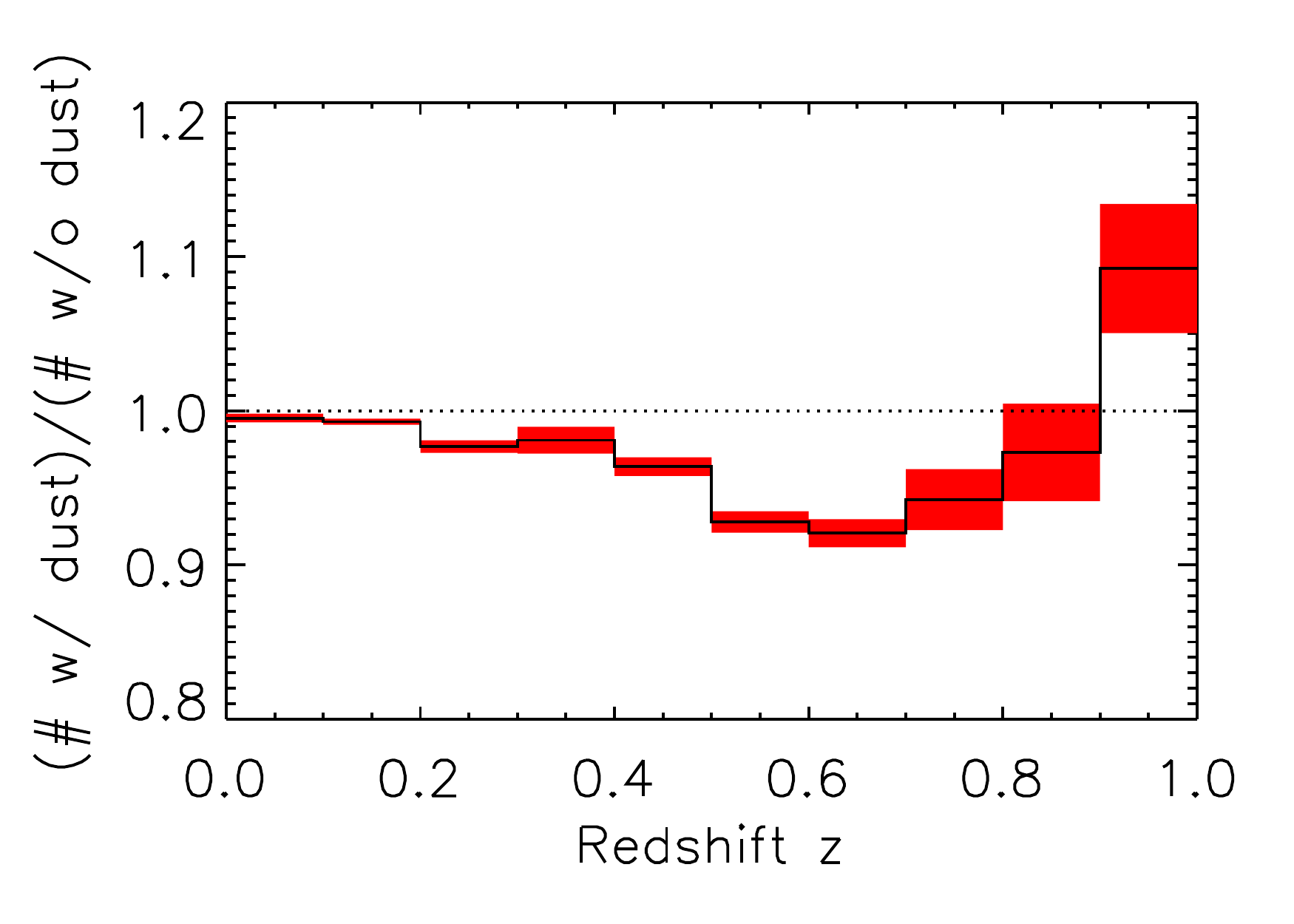}
\caption{Impact of dust emission on \Planck\ cluster completeness as a function of redshift.  The black line shows the completeness correction due to dust for the \Planck\ MMF cosmological catalog~\citep[S/N>6, \Planck\ 65\% cosmological mask,][]{planckszcosmo2016} computed from Monte Carlo simulations. The error bands in red are obtained from 10,000 bootstrap realizations.}
\label{fig:dust_completeness_correction}
\end{figure}

We compared the counts for recovered clusters with and without inclusion of dust emission. Results are shown in Fig.~\ref{fig:dust_completeness_correction}. The solid black line gives the total number of detected clusters with dust in the 10+10 simulations divided by the total number of detected clusters without dust, as a function of redshift. The red band corresponds to the standard deviation of 10,000 bootstraps over the 10+10 simulations. 

The dust emission significantly impacts the \Planck\ survey completeness over the redshift range [0.3 - 0.8], with a loss of $\sim 9\%$ of clusters in the [0.5-0.8] range. The [0 - 0.3] redshift range is only affected by $<2\%$ , as is the [0.8-0.9] bin. The [0.9-1] bin may present a small excess of detected clusters due to dust ($+9\%$), although the value is not statistically significant (the bootstrap error is 4.2\% in this bin).
We show in Appendix~\ref{app:cosmodep} that our dust model depends only weakly on cosmological parameters and, for simplicity, we adopt the curve shown in Fig.~\ref{fig:dust_completeness_correction} as the correction factor to apply to predicted counts before being associated with the observed counts in the \Planck\ likelihood.

\begin{figure*}
%\begin{figure}
\centering
\includegraphics[width=0.5\hsize]{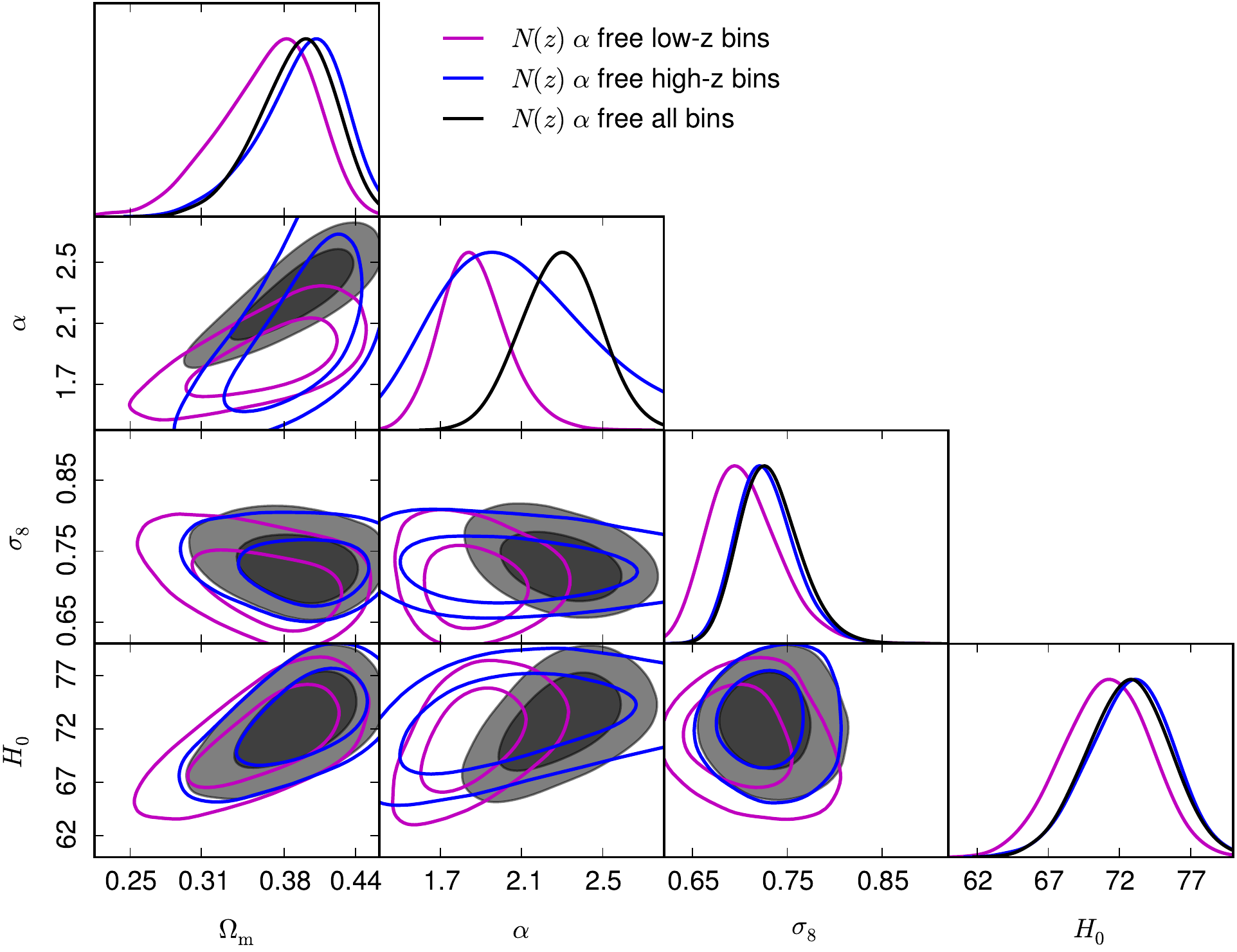}\includegraphics[width=0.5\hsize]{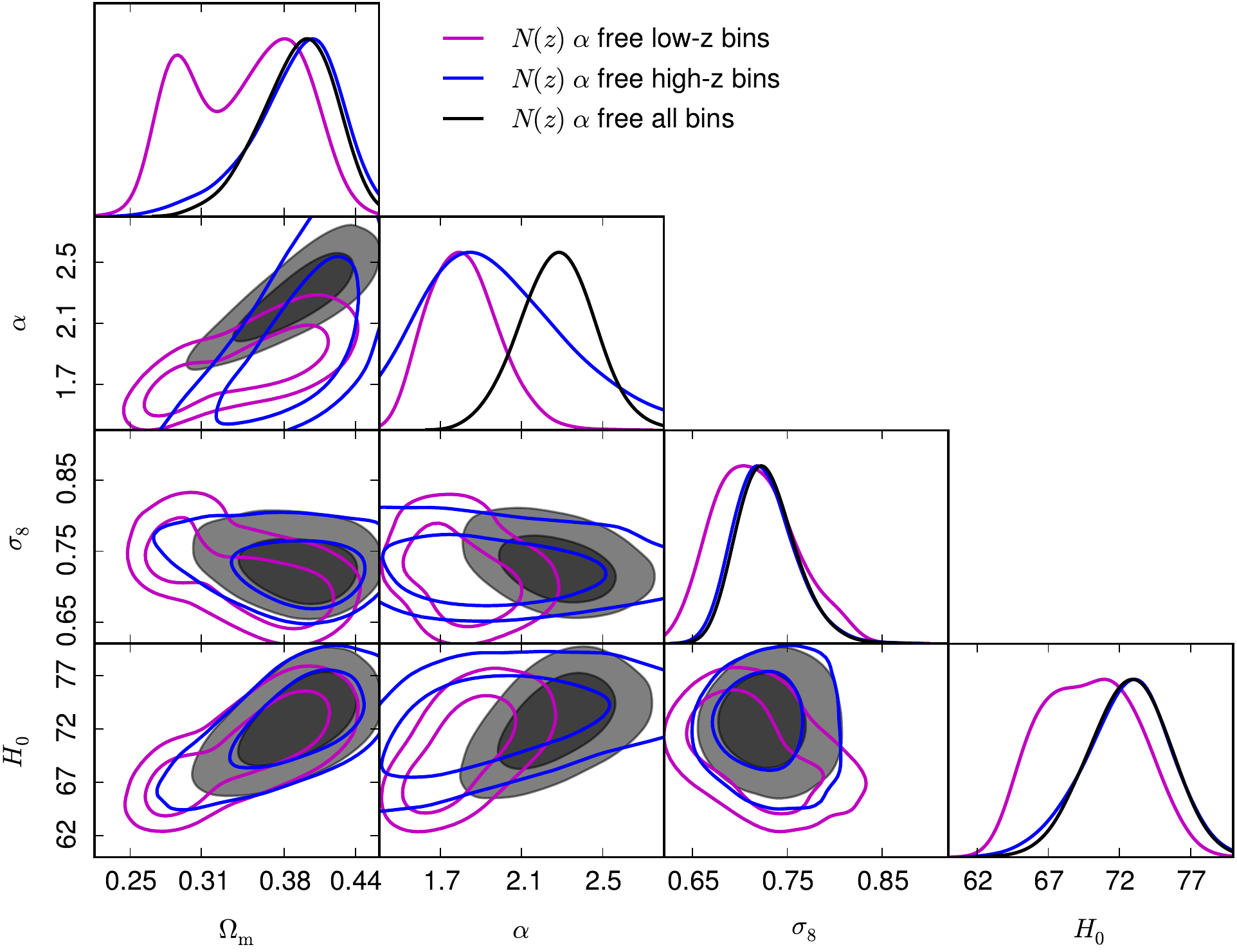}
\caption{{\it Left:} Cosmological parameters from the $N(z)$ \Planck\ likelihood when correcting the completeness for the effects of cluster dust emission (Fig~\ref{fig:dust_completeness_correction}). Shifts in cosmological parameters (black curves) are negligible with respect to the case when dust is not taken into account (right panel). {\it Right:} Cosmological parameters from the $N(z)$ \Planck\ likelihood without any dust correction to the completeness. This figure was obtained with the same likelihood as the original analysis~\citep[Fig.~7 of][]{planckszcosmo2016}, but the convergence of the chains is higher.}
\label{fig:cosmo_param_dust}
\end{figure*}

We reran the Markov chains for the $N(z)$ likelihood, correcting the completeness from the effects of the dust, over the full redshift range [0-1], and on the two distincts ranges [0-0.2] and [0.2-1], reproducing what was done in \cite{planckszcosmo2016}.  The results are shown in the lefthand panel of Fig.~\ref{fig:cosmo_param_dust}, presented in the same format as Fig.~7 of~\cite{planckszcosmo2016} to ease comparison. 

Over the entire redshift range [0-1] and also the high redshift range [0.2-1], the change in the contours is negligible. There is a change, on the other hand, for the low redshift range [0-0.2]; in particular, the $\Omega_{\rm m}$ posterior loses its bimodality.  This could be due to a lack of convergence of the original chains or to the fact that the low redshift likelihood is unstable. 

To decide between these two possibilities, we reran the original \Planck\ likelihood (i.e., without any dust correction to the completeness) to a higher level of convergence.  The results are shown in the righthand panel of Fig.~\ref{fig:cosmo_param_dust}. The bimodality of the $\Omega_{\rm m}$ posterior remains. We note that the low redshift contours in the righthand panel of Fig.~\ref{fig:cosmo_param_dust} and Fig.~7 of~\cite{planckszcosmo2016} differ slightly. Specifically, the maximum of the $\Omega_{\rm m}$ posterior is now the high $\Omega_{\rm m}$ solution, while it was the low $\Omega_{\rm m}$ solution in ~\cite{planckszcosmo2016}. We conclude that the low-$z$ likelihood is somewhat unstable. This is supported by the change in contour shape with increasing convergence, as just noted, and also by the fact that the dust correction in the first two redshift bins is small ($0.994$ for $0<z<0.1$ and $0.988$ for $0.1<z<0.2$). 

Despite this change in the low z likelihood, the two panels of Fig.~\ref{fig:cosmo_param_dust} are remarkably similar. This demonstrates that taking dust contamination into account in the analysis does not significantly change the preferred cosmological parameters, and does not ease the tension with the primary CMB.

%\end{figure}

\section{Conclusion}
\label{sec:ccl}

We have modeled dust emission in galaxy clusters at millimeter wavelengths using the model by~\cite{dezotti2016}, which we augmented by stacking PSZ2 clusters. The model now gives the shape of the dust profile and a normalization for the dust emission based on the \Planck\ 857~GHz channel. We used this model to simulate clusters that we injected into the \Planck\ maps.  We then assessed the impact of dust emission on \Planck\ cluster results, finding that:
\begin{itemize}
\item Dust emission is not responsible for the cluster size over-estimation seen in the real data.
\item The size over-estimation is plausibly caused by a mismatch, in the external regions, between the true cluster pressure profiles and the UPP adopted in the cluster extraction tool.
\item When fixing cluster size and position, dust emission biases \Planck\ cluster flux measurements low at only the 1 to 2\% level.
\item Dust emission impacts the completeness of the cluster cosmology catalog over the redshift range [0.3-0.8], with a maximum loss of $\sim 9\%$ of clusters between $z=0.5$ and $z=0.8$.
\item This cluster loss has a negligible effect on cosmological parameter estimation. Taking dust contamination into account in the \Planck\ cluster cosmology analysis does not help to ease the tension with the primary CMB.
\end{itemize}

Our constraints on the cluster dust emission model are general, and the calibrated model can be used to evaluate the impact of dust emission on other SZ surveys.  This is of particular interest for the highly sensitive next generation SZ cluster surveys such as the Advanced ACT \citep{advact}, SPT-3G \citep{spt-3g}, the Simons Observatory\footnote{\tt http://simonsobservatory.org}, CMB-S4 \citep{cmb-s4}, CORE~\citep{delabrouille2017,melin2017}, and PICO~\citep{hanany2018}.

\begin{acknowledgements}
The authors would like to thank the anonymous referee for useful comments that helped to clarify some important aspects of this work. JBM would like to thank G. Hurier for the detailed discussion about~\cite{hurier2016}. A portion of the research described in this paper was carried out at the Jet Propulsion Laboratory, California Institute of Technology, under a contract with the National Aeronautics and Space Administration. GDZ gratefully acknowledges financial support from ASI/INAF agreement n.~2014-024-R.1 and from the agreement ASI/Physics Department of the university of Roma--Tor Vergata n. 2016-24-H.0. Some of the results in this paper have been derived using the HEALPix~\citep{gorski2005} package.
\end{acknowledgements}

\appendix

\section{Comparison with published results}
\label{app:published_results}

In this Appendix, we compare our augmented~\cite{dezotti2016} model to previously published results. We discuss the comparison with the PSZ2 dust emission estimate from~\cite{hurier2016} in Appendix~\ref{app:hurier}, and comparison between the dust SED from~\cite{comis2016} with the SED from~\cite{cai2013} (used in the~\cite{dezotti2016} model) in Appendix~\ref{app:comis}.

\subsection{Comparison to PSZ2 dust emission from~\cite{hurier2016}}
\label{app:hurier}

\begin{figure}
\centering
\includegraphics[width=\hsize]{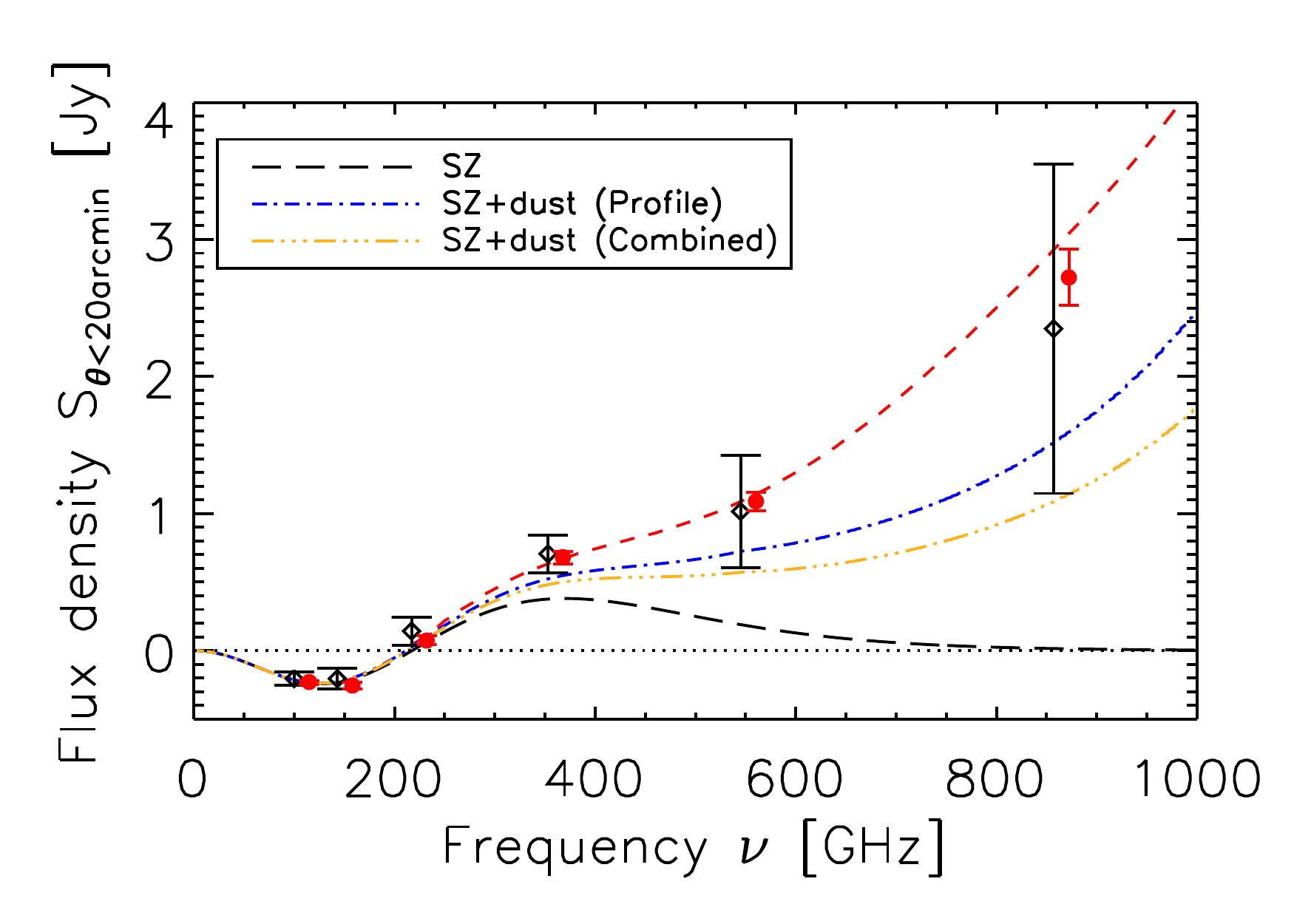}
\caption{Fixed aperture photometry. This figure is to be compared to Fig. 7 of~\cite{hurier2016}. The $y$-axis shows the average flux density of a cluster, so it needs to be multiplied by 1091 (the number of clusters in the analysis) to give the total flux density of the stack. Our data points are shown as black diamonds, and the points from~\cite{hurier2016} as red filled circles (shifted by +10~GHz for clarity). Our models are shown in blue and orange, and the model from~\cite{hurier2016} is shown in red. The black dashed line gives the contribution from the SZ signal alone.}
              \label{fig:stacked_flux_vs_nu}
\end{figure}

In Fig.~\ref{fig:stacked_flux_vs_nu}, we reproduce Fig. 7 from~\cite{hurier2016}. Instead of displaying the total flux density of the stack, we prefer to show the average value, so our $y$-axis must be multiplied by 1091(number of clusters in the analysis) to match the $y$-axis from ~\cite{hurier2016}. The central values (black diamonds) are in good agreement with the values found in~\cite{hurier2016} (displayed in our figure as red filled circles and shifted by +10~GHz for clarity), but our error bars are much larger. This could plausibly be due to the different methods used to estimate errors. 

We estimated our errors from the standard deviation of a bootstrap resampling of the sample of 1091 clusters, while \cite{hurier2016} computes the standard deviation at 1000 random locations on the sky. This second method does not capture the intrinsic variation of dust emission across the cluster population. Therefore, our fitted models (blue and orange lines) are fully consistent with our data points and error bars, but are significantly below the model proposed in~\cite{hurier2016} and shown as the red short dashed line. The black dashed line shows the contribution of the SZ signal only. 

The difference between Fig.~\ref{fig:mf_flux_vs_nu} and Fig.~\ref{fig:stacked_flux_vs_nu} comes from the difference in the measurement and averaging procedure. For Fig.~\ref{fig:mf_flux_vs_nu}, the signal extraction is performed on individual clusters within an area of radius $5 \times R_{500}$ using matched filters and assuming the profile from~\cite{arnaud2010}. With this template, the flux within $5 \times R_{500}$ is then converted to the flux within a sphere of radius $R_{500}$. Individual cluster fluxes are combined using an inverse-variance weighted average. For Fig.~\ref{fig:stacked_flux_vs_nu}, on the other hand, the signal is obtained from raw stacked maps and the error bars determined via bootstrap resampling. The flux in this case is estimated within a 20~arcmin radius aperture.

\subsection{Comparison of the SEDs from~\cite{comis2016} and~\cite{cai2013}}
\label{app:comis}

\begin{figure}
\centering
\includegraphics[width=\hsize]{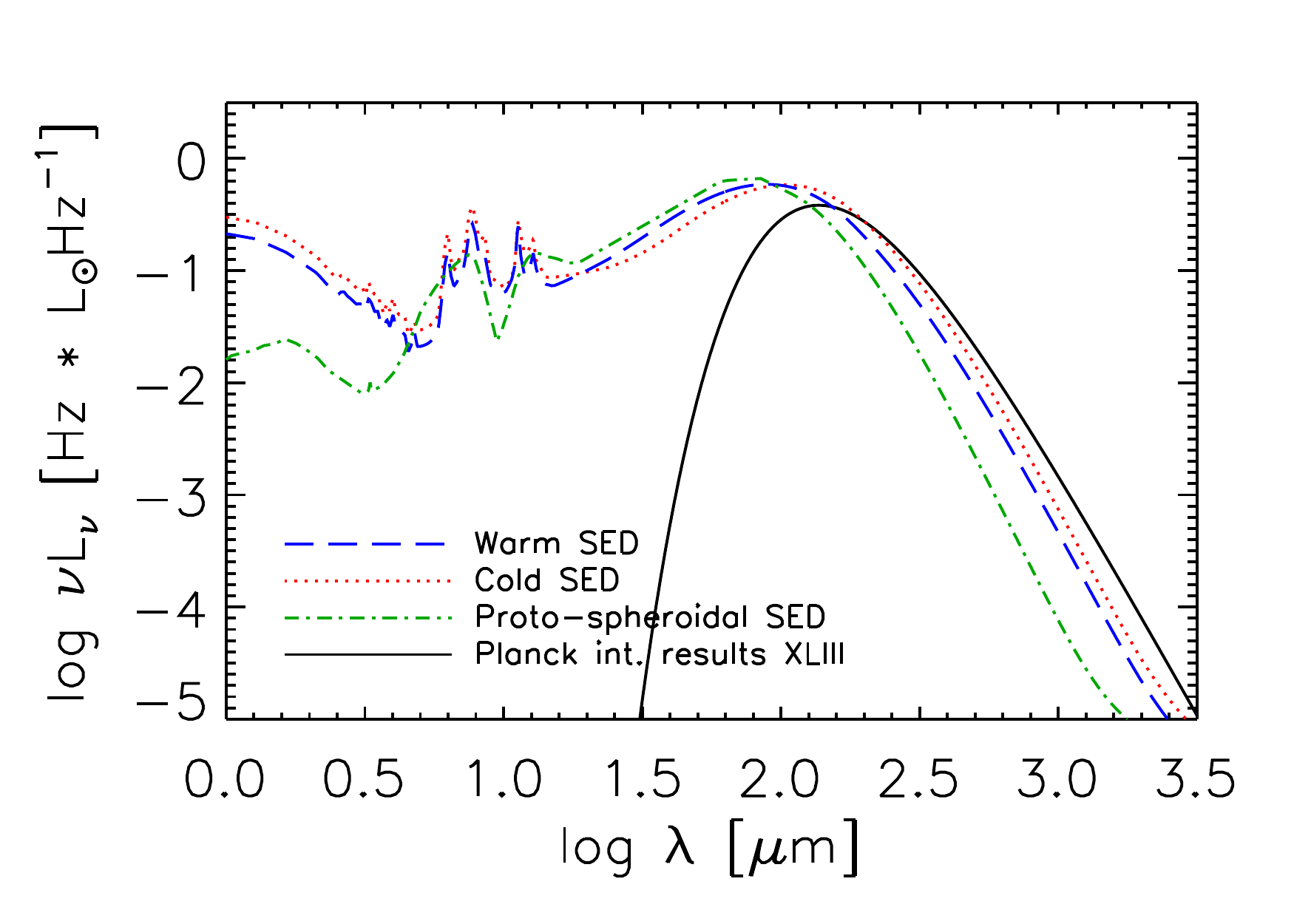}
\caption{Dust SEDs from Fig. 4 of~\cite{cai2013}.  We have added the best fit spectrum from~\cite{comis2016} as the solid black line.  It has been normalized so that the warm SED and the \Planck\ SED have the same integrated luminosity ($0.348 \, L_\sun$) between 100 and 3000~GHz (or equivalently between 100 and 3000~$\mu$m i.e., $\log \lambda$ between 2 and 3.48).}
              \label{fig:sed_cai_vs_planck}
\end{figure}

In Fig.~\ref{fig:sed_cai_vs_planck}, we compare the SED determined on PSZ2 clusters in~\cite{comis2016} to the SED from~\cite{cai2013} used in the~\cite{dezotti2016} model.  The frequency dependence of the warm and cold SEDs of~\cite{cai2013} are similar to the SED from the PSZ2 over the range covered by \Planck\ (100 to 857~GHz, or equivalently from 350 to 3000~$\mu$m i.e., $\log \lambda$ between 2.54 and 3.48). This explains why our model, although only adjusted to the 857~GHz data, also provides a good match at lower \Planck\ frequencies in Fig.~\ref{fig:mf_flux_vs_nu}.

\section{Blind flux estimation of PSZ2 clusters}
\label{app:psz2_flux}

\begin{figure*}
%\begin{figure}
\centering
\includegraphics[width=0.45\hsize]{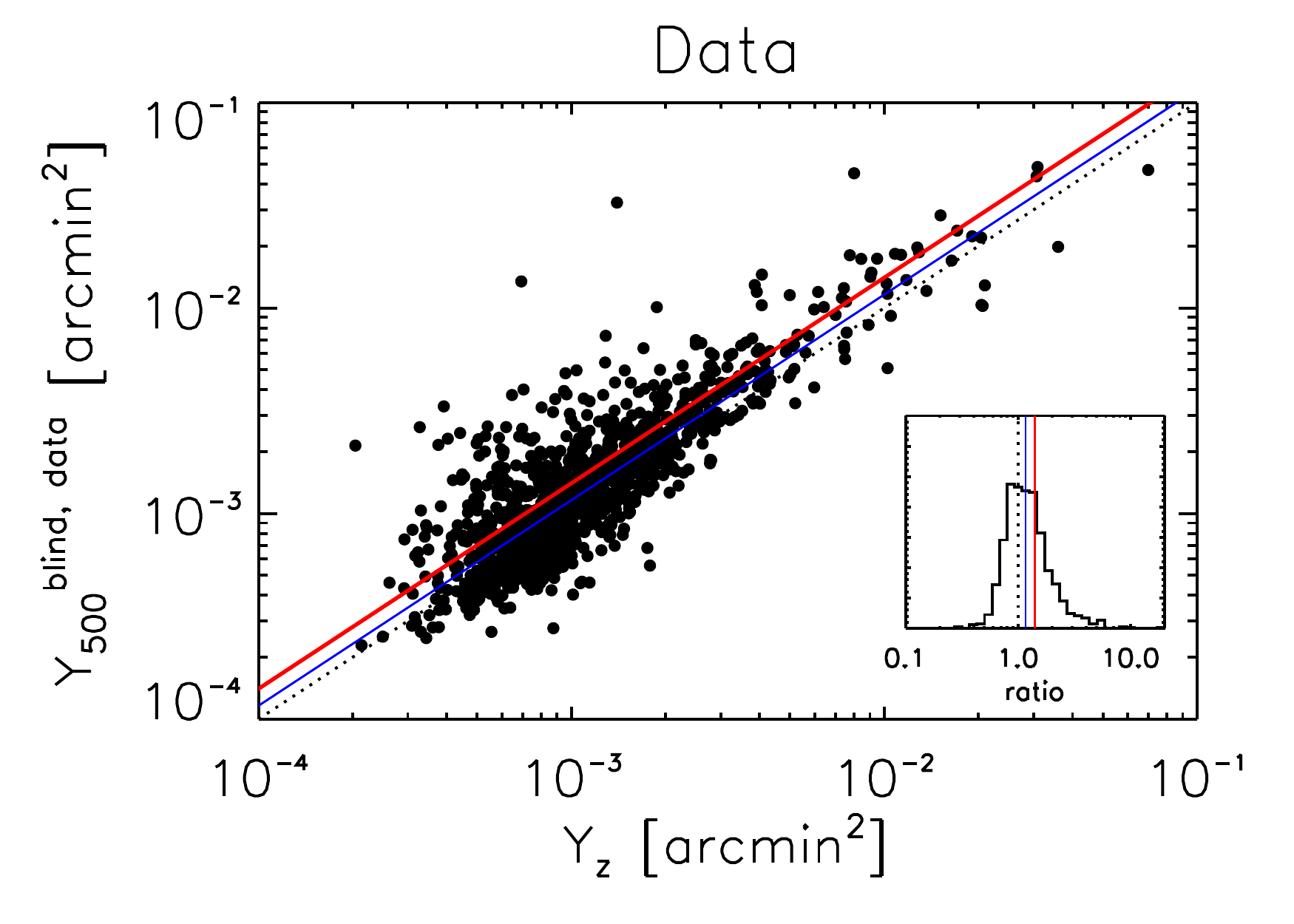} \includegraphics[width=0.45\hsize]{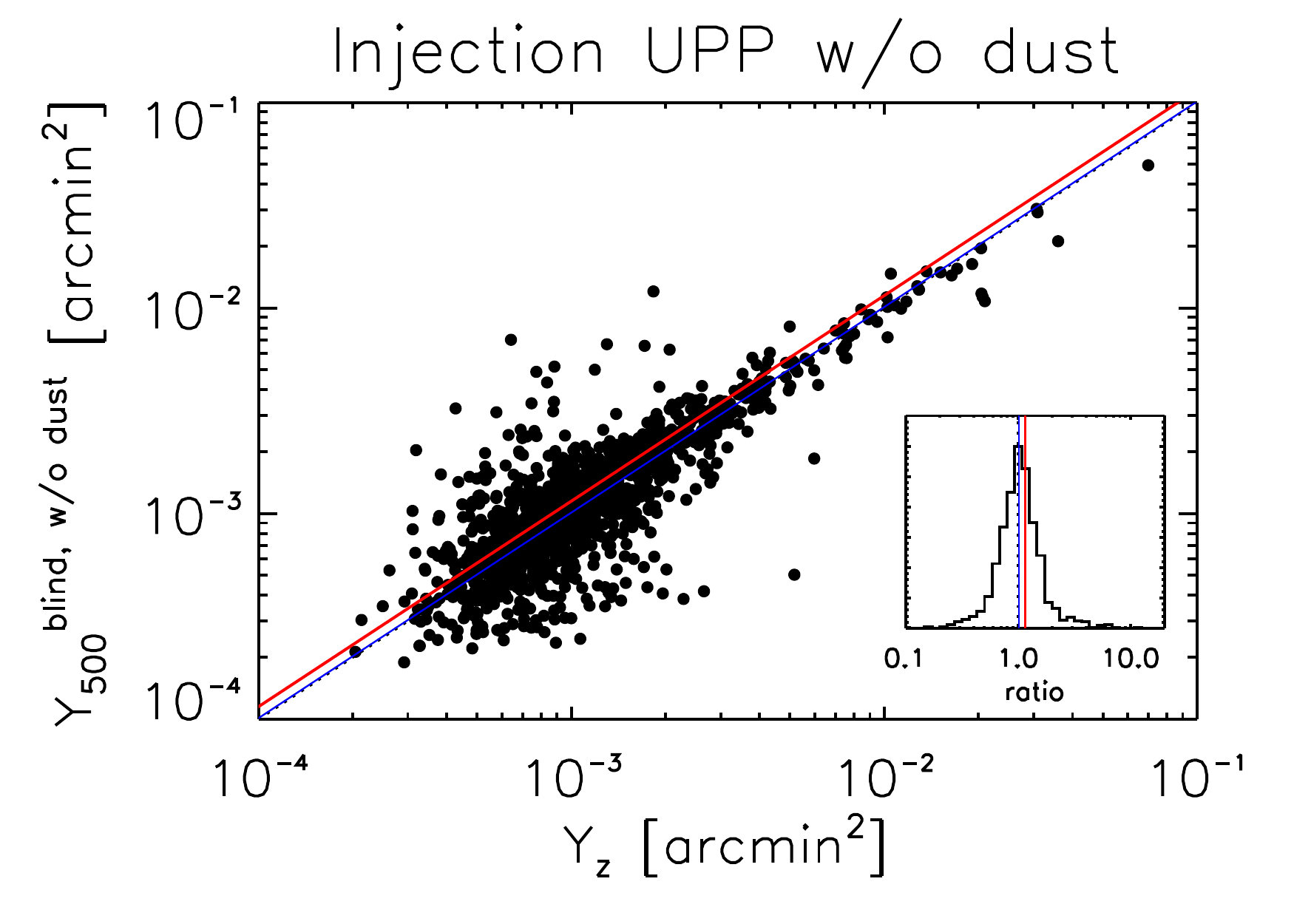}
\includegraphics[width=0.45\hsize]{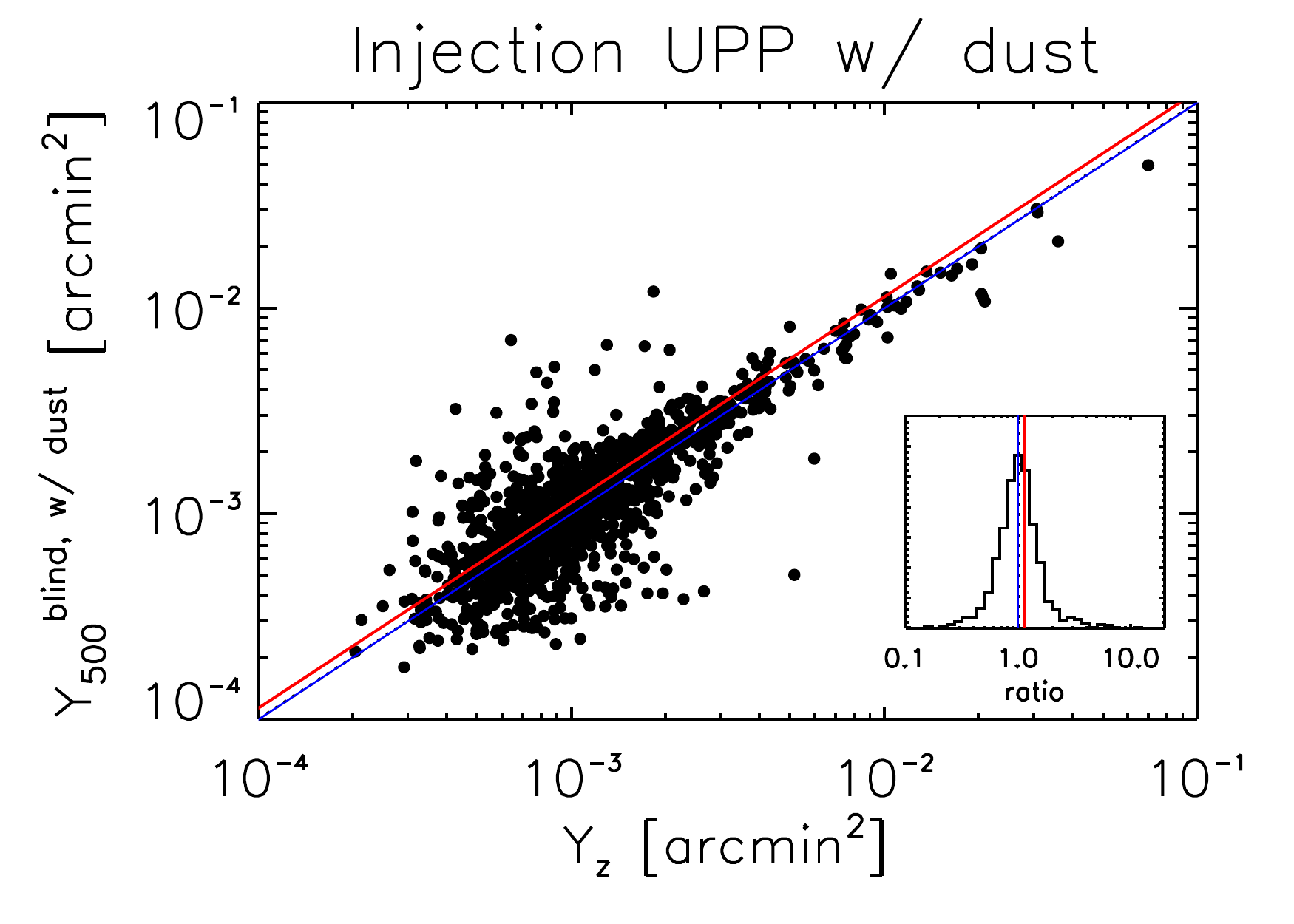} \includegraphics[width=0.45\hsize]{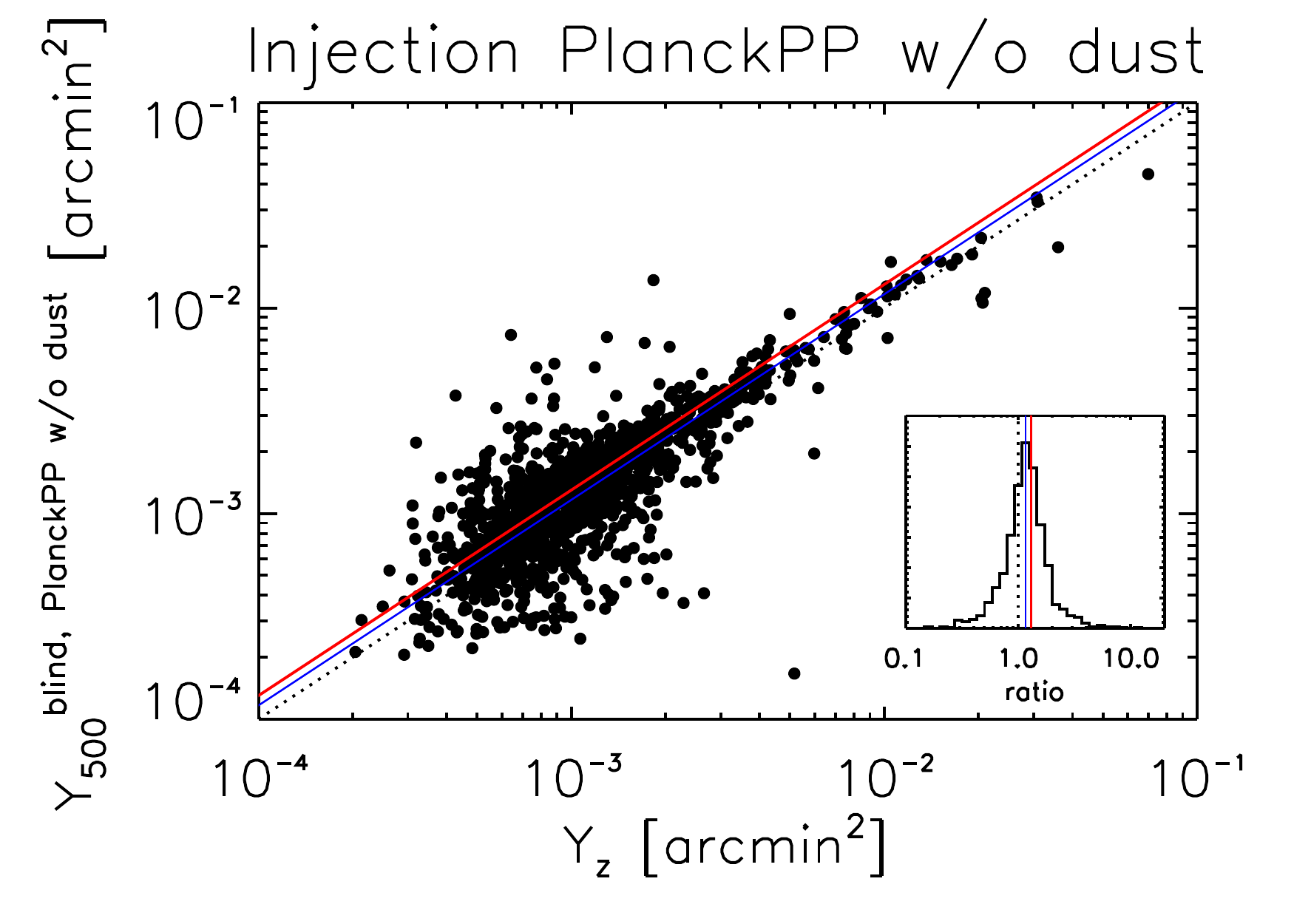}
\caption{Same as Fig.~\ref{fig:recovered_sizes}, but for the flux $Y_{500}^{\rm blind}$ (extracted blindly) and $Y_z$ (from the \Planck\ mass proxy). {\it Top left:} For the data, the mean (median) of the ratio $Y_{500}^{\rm blind} \over {Y_z}$ is 1.40 (1.16) and is displayed as the red (blue) line. The thickness of the line encapsulates the 68\% error on the mean (median) calculated with a bootstrap. The histogram of this ratio is shown in the inset. {\it Top right:} For the injection of PSZ2 clusters using the UPP and without the dust component; the mean (median) is 1.15 (1.01). {\it Bottom left:} For the injection of PSZ2 clusters using the UPP and with the dust component; the mean (median) is 1.13 (1.00). {\it Bottom right:} For the injection of PSZ2 clusters using the PlanckPP and without the dust component; the mean (median) is 1.30 (1.17) - close to the values found with the data. The dotted line in all four panels is the equality line.}
              \label{fig:recovered_fluxes}
\end{figure*}

In this Appendix, we compare the SZ signal extracted blindly to the injected signal for simulated PSZ2 clusters. Fig.~\ref{fig:recovered_fluxes} is equivalent to Fig.~\ref{fig:recovered_sizes}, but for the flux instead of the size. The figure shows that the over-estimation of the blind flux is not due to the dust emission in clusters. It likely finds its origin in the mismatch between the profile assumed for cluster extraction (UPP) and the actual cluster profile (closer to the PlanckPP), as already shown in Fig.~\ref{fig:recovered_sizes} for cluster size and discussed in Sect.~\ref{sec:clusterfluxsize}.

%\end{figure}

\section{Stacked maps}

\begin{figure*}
%\begin{figure}
\centering
\includegraphics[width=0.33\hsize]{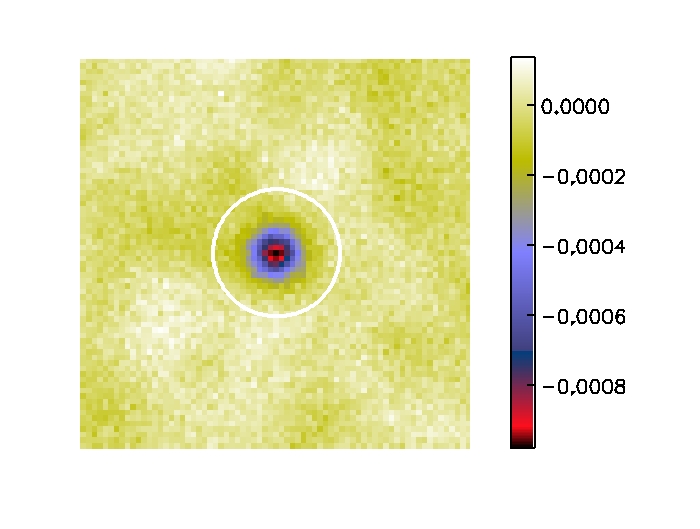} \includegraphics[width=0.33\hsize]{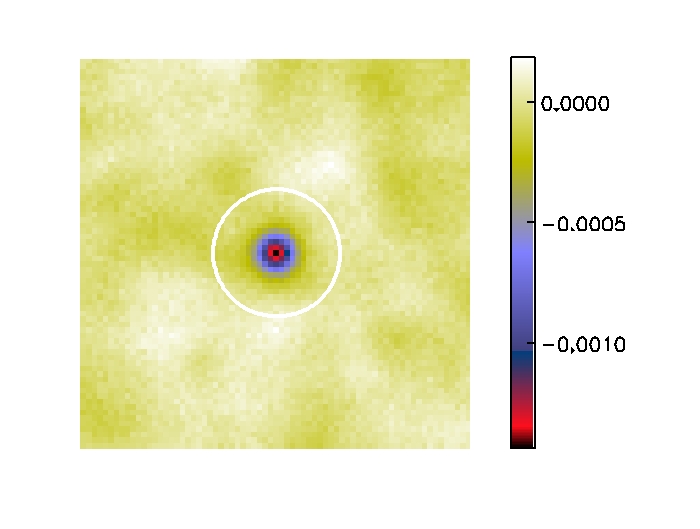} \includegraphics[width=0.33\hsize]{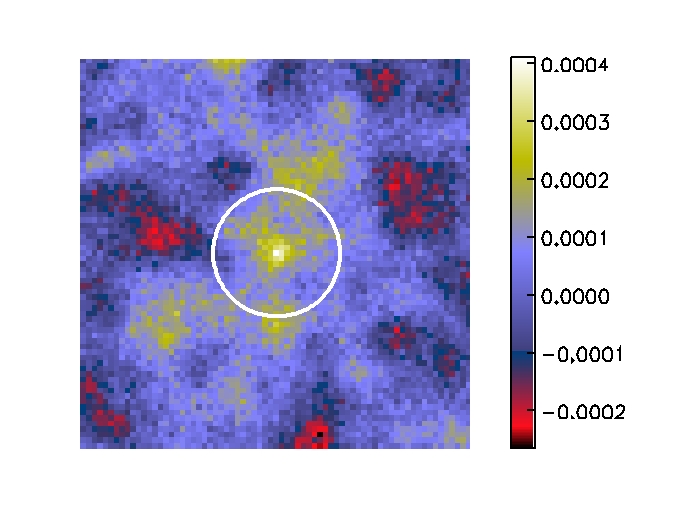} \\
\includegraphics[width=0.33\hsize]{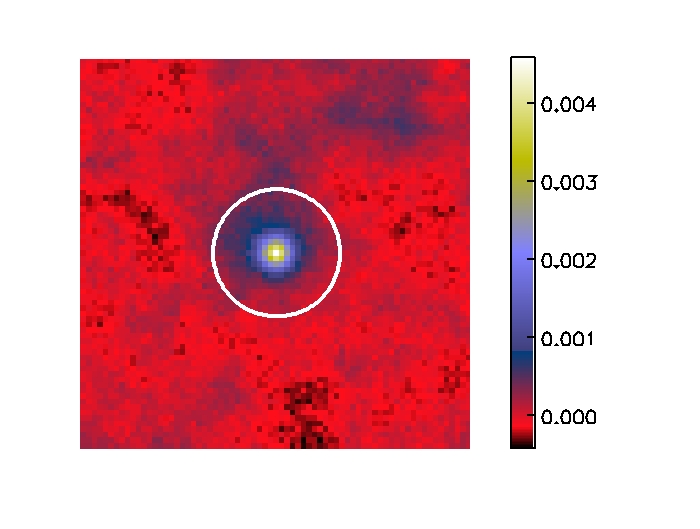} \includegraphics[width=0.33\hsize]{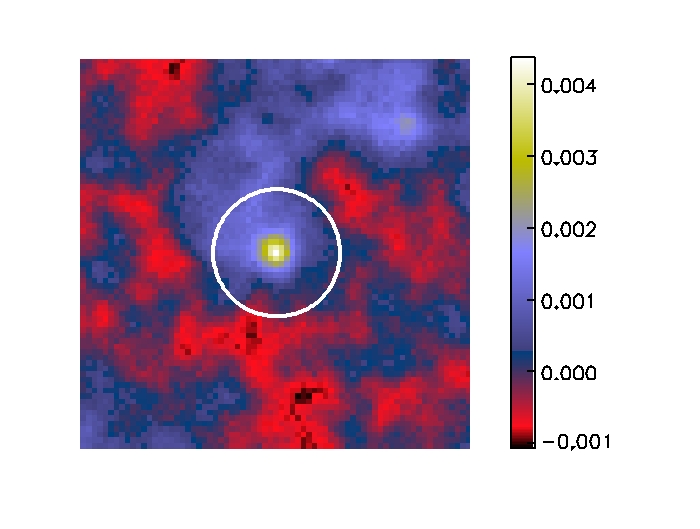} \includegraphics[width=0.33\hsize]{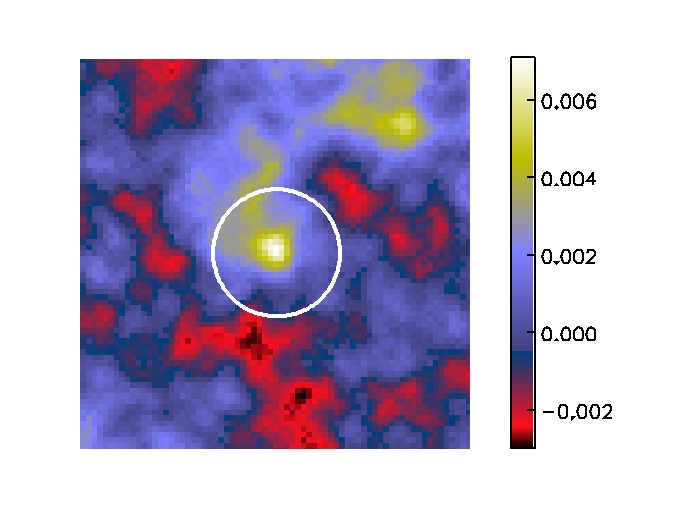} 
\caption{{\it From left to right and top to bottom:}  \Planck\ HFI maps at 100, 143, 217, 353, 545 and 857~GHz stacked on cluster positions. The maps are $2 \times 2 \rm \deg^2$ in units of ${\rm Jy/arcmin}^2$. One can clearly see the decrement of the SZ effect at 100 and 143~GHz. The dust emission, mixed with the increment of the SZ emission, is visible at higher frequencies. The white circle is centered on the stack position and is 20~arcmin in radius. The central values (black diamonds) and associated error bars of Fig.~\ref{fig:stacked_flux_vs_nu} are obtained as the integral of the signal from these maps within the white circles.}
              \label{fig:stacked_maps_on_clusters}
\end{figure*}

\begin{figure*}
%\begin{figure}
\centering
\includegraphics[width=0.33\hsize]{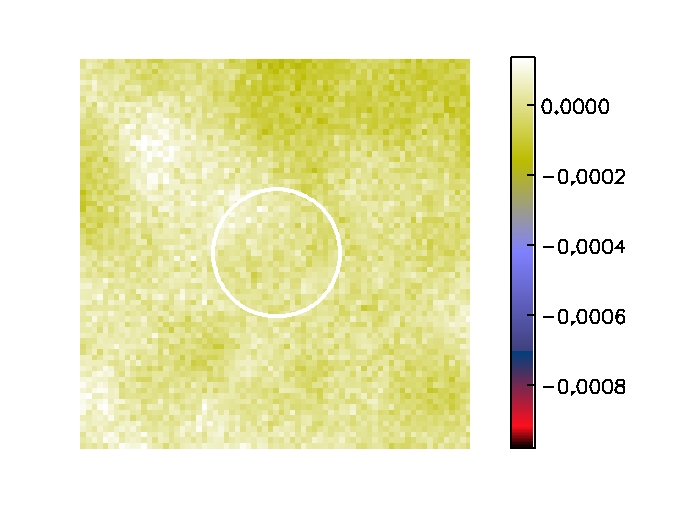} \includegraphics[width=0.33\hsize]{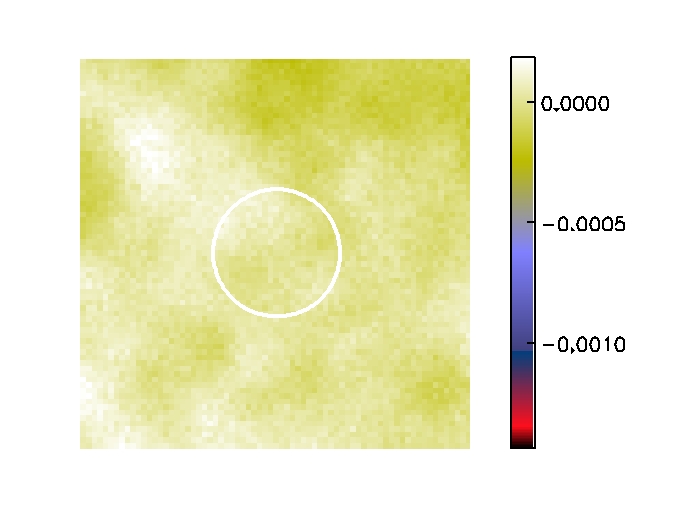} \includegraphics[width=0.33\hsize]{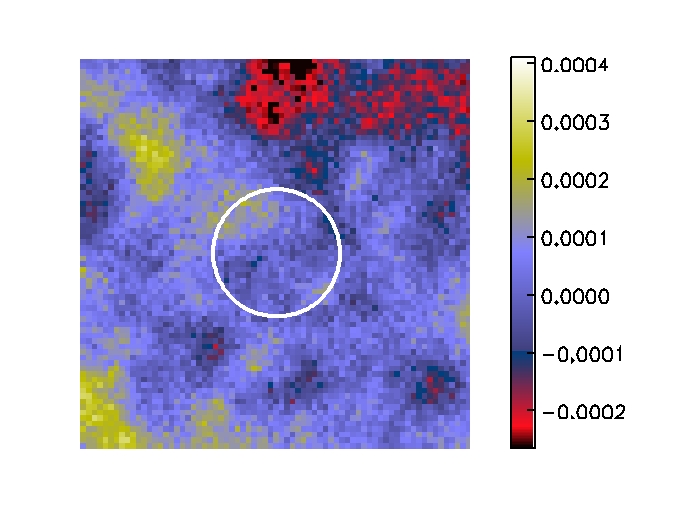} \\
\includegraphics[width=0.33\hsize]{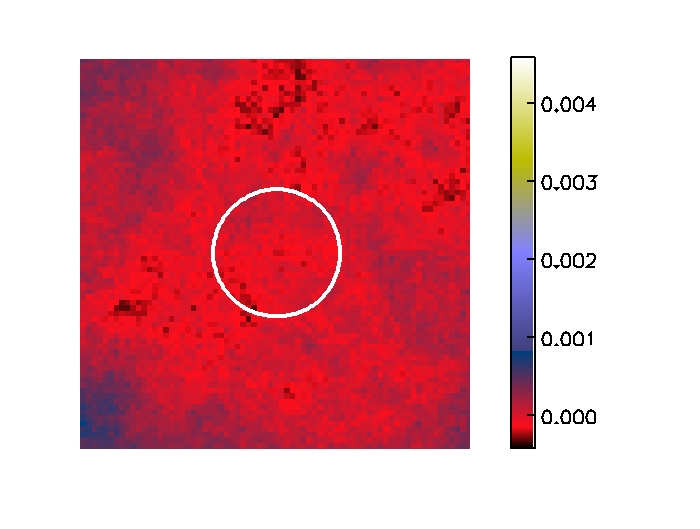} \includegraphics[width=0.33\hsize]{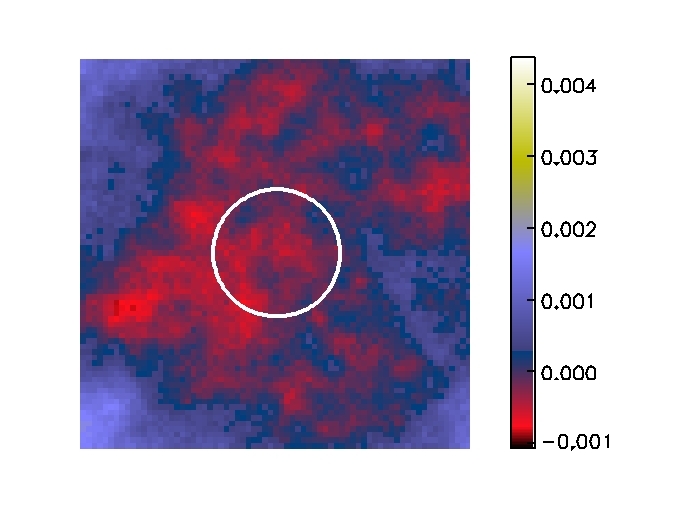} \includegraphics[width=0.33\hsize]{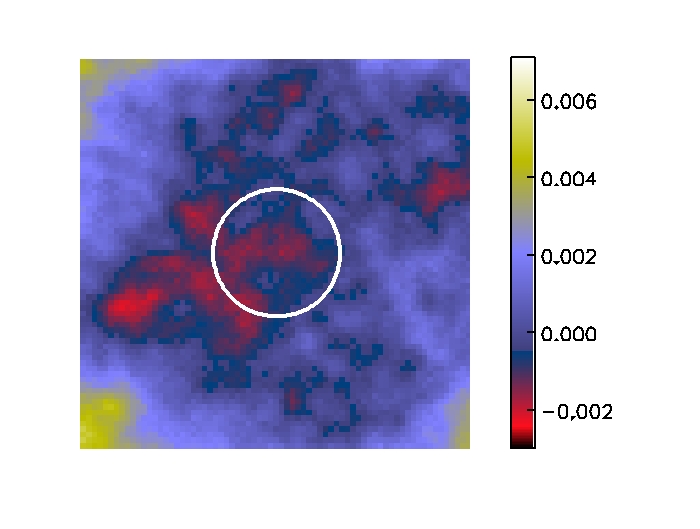} 
\caption{{\it From left to right and top to bottom:}  \Planck\ HFI maps at 100, 143, 217, 353, 545 and 857~GHz stacked at random positions. The maps are $2 \times 2 \rm \deg^2$ in units of ${\rm Jy/arcmin}^2$. We have adopted the same color scales as in Fig.~\ref{fig:stacked_maps_on_clusters}. The white circle is located at the stack center and is 20~arcmin in radius.}
              \label{fig:stacked_maps_off_clusters}
\end{figure*}

We show \Planck\ maps stacked at the PSZ2~(Fig.~\ref{fig:stacked_maps_on_clusters}) positions and at random positions~(Fig.~\ref{fig:stacked_maps_off_clusters}). In Fig.~\ref{fig:stacked_maps_on_clusters}, the negative part of the SZ effect is clearly visible at 100 and 143~GHz, and the dust contribution mixed with the increment of the SZ emission is visible at frequencies above 217~GHz. No significant emission is found in Fig.~\ref{fig:stacked_maps_off_clusters}.

%\end{figure}

%\end{figure}

\section{Cosmology dependence of the effect of dust emission on \Planck\ completeness}
\label{app:cosmodep}

The effect of dust emission on the \Planck\ completeness may depend on cosmology, in particular because the dust model fit performed in Sect.~\ref{sec:constraints} may depend on the assumed cosmological parameters. Expressing the completeness as a function of redshift introduces an additional dependance on cosmology. It is technically feasible to express it as a function of cluster flux and size, as in the \Planck\ analyses, to avoid this latter cosmological dependance, but this would require significant additional computing time to run more simulations to build the two dimensional quantity. The major difficulty would be to assess the  dependence on cosmology: we would need to Monte Carlo the whole analysis chain (fit for the dust model, simulations, injections, extractions) on each set of cosmological parameters, which would require some unmanageable computing time. Running the full analysis takes about two weeks for a single cosmology.

\begin{figure}
\centering
\includegraphics[width=\hsize]{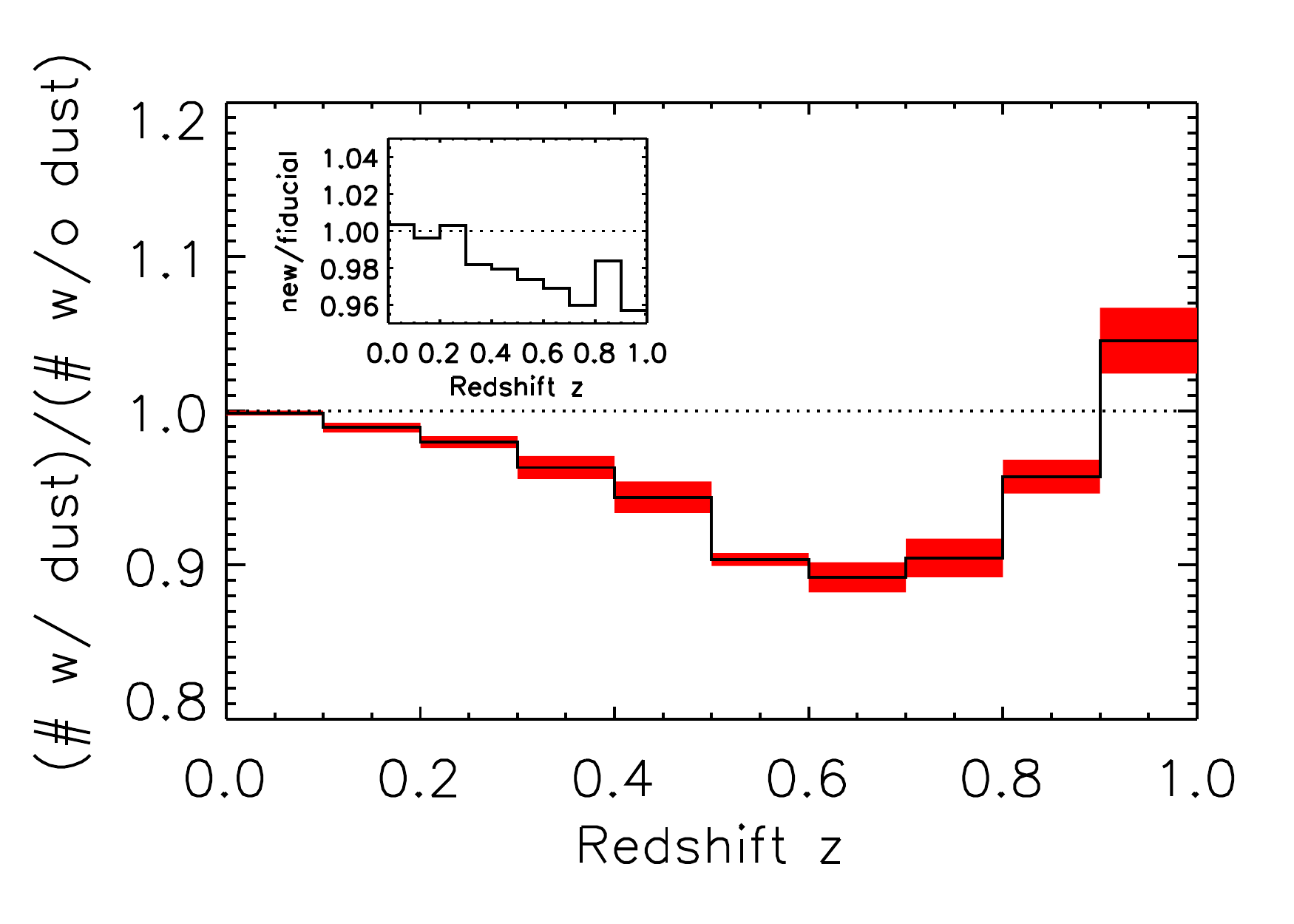}
\caption{Impact of dust emission on \Planck\ cluster completeness as a function of redshift for a flat $\Lambda$CDM model with $\Omega_{\rm m}=0.4$ and $\Omega_\Lambda=0.6$.  The black line shows the completeness correction due to dust for the \Planck\ MMF cosmological catalog~\citep[S/N>6, \Planck\ 65\% cosmological mask,][]{planckszcosmo2016} computed from Monte Carlo simulations. The error bands in red are obtained from 10,000 bootstrap realizations. The inset shows the ratio between the black line of this figure and the one from Fig.~\ref{fig:dust_completeness_correction}.}
\label{fig:dust_completeness_correction_othercosmo}
\end{figure}

In order to test the cosmology dependance of the effect of dust on \Planck\ completeness, we thus performed a second full analysis and changed the value for $\Omega_{\rm m}$ to 0.4 and $\Omega_\Lambda$ to 0.6, while keeping the other parameters fixed to the \Planck\ $\Lambda$CDM cosmology. This model is located far from our fiducial \Planck\ $\Lambda$CDM cosmology in the 95\%C.L. region of the \Planck\ cluster cosmological constraints~\citep[Fig. 7 of][]{planckszcosmo2016}. The impact of the adopted model on completeness is shown in Fig.~\ref{fig:dust_completeness_correction_othercosmo} and the ratio between the black line of this figure and the one from Fig.~\ref{fig:dust_completeness_correction} is shown in the inset. The change in the effect of dust emission between the two sets of cosmological parameters is weak ($<4\%$ in the full \Planck\ cluster redshift range $[0-1]$ as shown in inset). We thus adopted the curve from Fig.~\ref{fig:dust_completeness_correction} to correct our predicted cluster counts for all sets of cosmological parameters.

\bibliographystyle{aa}
\bibliography{dustinclusters}

\end{document}